# '*Marrying Up*' or '*Matching Even*'? Socioeconomic Drivers of Spousal Age Gaps in India

**Praveen**[1],

**Siddhanta S.**[2],

**and Chaudhuri A.**[3]

E-mail address:

a. Praveen
praveenjai.jk@gmail.com

b. Suddhasil Siddhanta
suddhasil.siddhanta@gipe.ac.in

c. Anoshua Chaudhuri
anoshua@sfsu.edu

[1] Center of Policy Research and Governance, New Delhi, India

[2] Gokhale Institute of politics and Economics, Pune, Maharashtra, India

[3] Department of Economics, San Francisco State University, CA, USA

**Abstract:** This study examines the determinants of the spousal age gap (SAG) in India, utilizing data from the 61st and 68th rounds of the National Sample Survey (NSSO). We employ regression analysis, including instrumental variables, to address selection bias and account for unobservable factors. We hypothesize an inverted U-shaped relationship between educational assortative mating and SAG, where, keeping the husband's education constant at the graduation level, the SAG first widens and then narrows as the wife's education level increases from primary to postgraduate. This pattern is shaped by distinct socio-economic factors across rural and urban contexts. In rural areas, increasing prosperity, changes in family structure, and educational hypergamy contribute to a wider age gap, with the influence of bride squeeze further exacerbating this disparity. Conversely, in urban areas, while the growth of white-collar jobs initially contributed to a narrowing of the SAG in 2004-05, this trend did not persist by 2011-12. Specifically, the influence of income on SAG becomes nonlinear, showing declining trends beyond the 7th income quantile, reflecting limited marriage mobility opportunities for females and hinting at a possible threat to the institution of marriage among the urban upper class. To the best of our knowledge, this is the first study to provide empirical evidence on how specific social, economic, and cultural dynamics influence the spousal age gap in Indian society. This increasing or persistent spousal age gap has significant implications for the treatment of women, power dynamics, and violence within marriage.

## 1. Introduction

The traditional societal norm of the "male-breadwinner and female-homemaker" model has rendered age hypergamy—wherein husbands are older than their wives—a near-universal feature of marital dynamics (Bergstrom and Bagnoli, 1993) worldwide. Age hypergamy confers benefits on men, as marrying younger wives is associated with increased male longevity, but it is detrimental for women: marrying an older husband reduces female lifespan, though marrying a younger one shortens it even further. In contrast, age homogamy optimizes longevity for women (Drefahl, 2010).

Age hypergamy is particularly prevalent in patriarchal societies where traditional gender roles are strictly enforced. While India exhibits this stalling trend, similar patterns emerge across developing economies. Studies across Sub-Saharan Africa document declines of 1.7 years between 1980-2014, yet spousal age gaps persist at 4.5-12.9 years across countries (Hirschi et al., 2024; Kyei et al., 2025). Cross-national analyses further show that economic development inversely correlates with these gaps, moderated by cultural and institutional factors (Casterline et al., 1986; Esteve et al., 2012).

In India, husbands are, on average, four to five years older than their wives. Data from 1991 to 2011 reveal a stalling trend with only a marginal decline in the spousal age gap (SAG) nationwide (from 4.7 to 4.4 years in rural areas and from 4.9 to 4.6 years in urban areas). However, regional variations exist: the SAG has increased in the southern state of Kerala, rural Tamil Nadu, rural Odisha, and rural Himachal Pradesh[4].

**Figure 1a and 1b**: Spousal Age Gap by Monthly Per Capita Consumption Expenditure, NSS 50th (1993-94), 61st (2004-05) and 68th (2011-12) rounds, All India

1(a) Rural Sector 1(b) Urban Sector

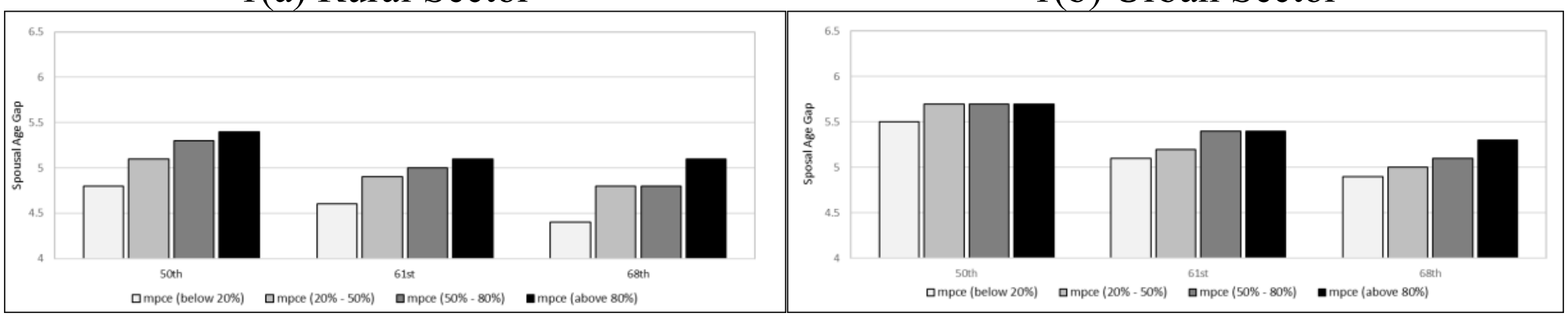

Economic development often reinforces these entrenched gender roles. India's low and declining female labor force participation amid rising husband incomes (Datta Gupta et al., 2020), ongoing masculinization of the sex ratio (Agnihotri, 2000; Siddhanta et al., 2003), and the diffusion of gender norms across cultural communities (Razavi and Miller, 1995) indicate persistent gender disparities despite socioeconomic progress. This aligns with broader evidence from marriage market studies, which highlight how spousal sorting is influenced by labor opportunities and cultural norms, with downstream effects on gender wage gaps and household inequality (Calvo et al., 2021; Eika et al., 2019). Data from successive Employment and Unemployment Surveys by the National Sample Survey Organisation (NSSO), Government of India, reflect a similar pattern: the SAG widens with increasing prosperity (see Figures 1a and 1b)[5].

---

[4] For regional variations, see Table 1 in Appendix 1

[5] Regional pattern also supports national patterns, but with notable variations. In 2011-12, the Spousal Age Gap (SAG) for the highest income group stood at 5.1 years in rural and 5.2 years in urban sectors. However, the gap

Education, however, acts as a taste-shifter. Beyond a threshold (typically middle or secondary level), it enhances women's access to better-paying jobs and raises their opportunity costs. Holding the husband's education fixed at the graduate level, the relationship between the SAG and the educational gap follows an inverted-U pattern as women's education advances from illiterate to higher levels (Figures 2a and 2b). The SAG is narrower at the extremes—when wives are illiterate or primary-educated, or when they hold graduate degrees or beyond—and widest in the intermediate range (middle, secondary, or higher secondary education)[6].

This evidence underscores how socioeconomic factors shape the SAG: a husband's higher income and educational hypergamy may exacerbate age hypergamy, whereas women's advanced education tends to mitigate it. We propose a conceptual framework linking the SAG to sectoral transformations and utilize Indian microdata from three NSSO waves—the 50th round (July 1993–June 1994), the 61st round (July 2004–June 2005), and the 68th round (July 2011–June 2012)—to model its determinants. Employing instrumental variable techniques to address selection bias and unobserved heterogeneity, our analysis demonstrates that communities with pronounced educational hypergamy exhibit greater age hypergamy, while educational homogamy and hypogamy can counteract it.

**Figure 2a and 2b**: Spousal Age Gap by Spousal Education Gap, NSS 50$^{th}$ (1993-94), 61$^{st}$ (2004-05) and 68$^{th}$ (2011-12) rounds, All India, Urban Sector

2 (a) Rural Sector 2 (b) Urban Sector

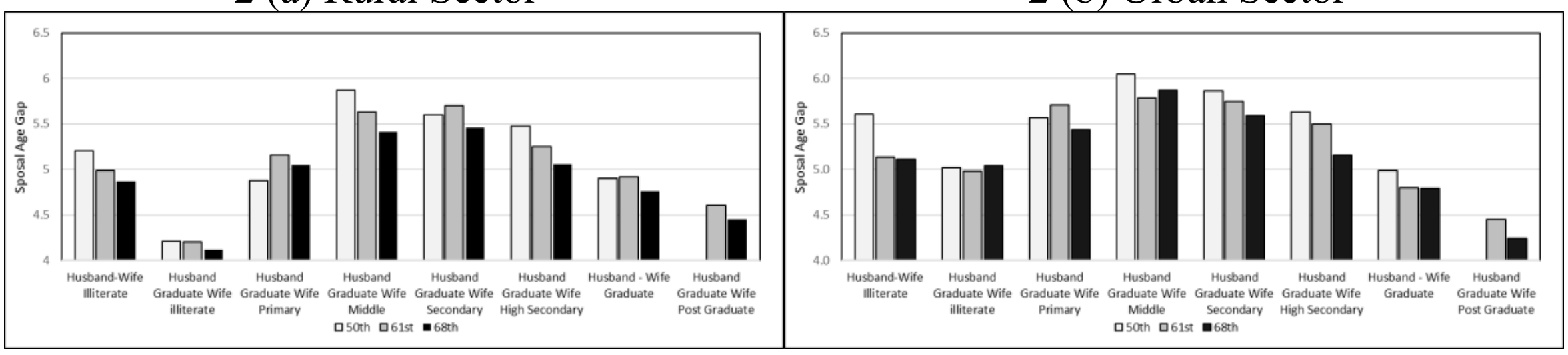

This study contributes through an in-depth examination of socioeconomic influences on spousal age differences. Extending beyond descriptive trends, it investigates the interplay of education and income in these dynamics. By exploring how women's education disrupts traditional age hypergamy, it contributes to the literature on gender inequality, marriage markets, and economic development in patriarchal societies. Additionally, it offers fresh empirical evidence on educational transitions' effects on marital outcomes, enriching discussions on gender and social mobility. These findings have important policy implications for interventions targeting gender equality, education access, and women's economic was wider in southern India (6.4 years in rural and 6.0 years in urban sectors). Eastern India's wealthiest bracket exhibited the largest gap at 6.8 years in 2011-12, (though it was 7.1 years in 1993-94). Thus, in spite of some longitudinal correction, increased prosperity consistently correlates with a larger SAG (see Appendix 2) suggesting that wealthier men tend to marry younger women, leading to a 'gender penalty' for women.

[6] At more disaggregate level, the southern India tends to have a larger SAG in response to education-hypergamy. The central region, on the other hand, maintains a relatively stable age gap across different educational levels of wives, while the rural sector of the Eastern region exhibits the widest age gap when wives have primary to secondary levels of education. However, this gap drops significantly when wife attains a graduate and above levels of education. Despite identical education levels and different regional affiliations, the urban sector consistently displays a larger SAG compared to its rural counterpart. This suggests that traditional tendencies may be more pronounced in ostensibly modern (urban) settings.

empowerment, as reducing spousal age gaps can enhance women's autonomy and decision-making power within households (Kyei and Bawah., 2024).

The paper proceeds as follows: Section 2 discusses the theory behind an inverted 'U' shaped age hypergamy hypothesis, and the role of income & substitution effects, and stigma and gender egalitarian norms in the marriage and labor markets. Section 3 describes the data. Section 4 details indices of economic transformation and household transition, and explains the formulation of our stigma and gender-egalitarian norms index. Section 5 presents empirical models and results, and Section 6 is the conclusion and discussion.

## 2. Inverted U Hypothesis in Spousal Age Gap

The persistence and variability of the spousal age gap (SAG) in different societies can be traced to the gendered evolution of social norms. In traditional settings characterized by agricultural economies and limited returns to education, a man's desirability in the marriage market is closely tied to his capacity to prove his earning potential over time. By contrast, a woman's marital "value" has historically been associated with physical maturity and childbearing potential, leading to earlier marriages for women and resulting in larger age gaps when average marriage age is low (Figure 3).

As societies develop and educational attainment rises, shifts occur in the patterns of age-hypergamy. In the early stages of structural transformation—from agriculture to manufacturing—men initially benefit due to their comparative advantage in physically demanding work (Alesina et al., 2013). Meanwhile, deeply embedded gender norms and labor market stigmas often restrict women's opportunities, especially in manufacturing, where a wife's employment may be perceived as reflecting the husband's inability to provide (Goldin, 1994). This effect deepens in regions facing a shortage of marriageable women – 'bride deficits', these dynamics are often exaggerated: the marriage market rewards men's earning power while stigma limits female labor force participation, further amplifying the spousal age gap and reinforcing hypergamy (Figure 3).

The transition to services and rising demand for white-collar jobs can equalize marriage timing for both genders, driven by increasing opportunity costs. Social stigma against wives in white-collar roles is lower, as "highly educated women across many cultures are given license to work for pay" (Goldin, 1995), potentially compressing age gaps through educational homogamy (Figure 3). This holds especially if partners favor similar traits (matching hypothesis; Kalmijn, 1994; Schwartz, 2013). Yet, in societies with significant "missing brides," income effects may surge if traditional female roles are highly valued. Even then, educational homogamy can reduce gaps if partners compete for superior characteristics (competition hypothesis), avoiding "marrying down" and yielding comparable pairings (Chiappori et al., 2017).

As women specialize in their fields and surpass men in scholastic achievement, they enter the marriage market later than many male peers from the same cohort. This delay imposes opportunity costs but positions them as informed, educated individuals beyond homemaking roles. When technology substitutes unpaid labor, modern contraceptives gain acceptance, paid work faces less sanction (Goldin, 1995), or societies view female education and participation as modernity markers (Allendorf & Pandian, 2016), highly educated women can prioritize personal traits in mate selection—narrowing age gaps (Figure 3)—provided mutual agreement on trait matching. Although class shapes lifestyles, attitudes, and beliefs (Weeden & Grusky, 2012), advanced female education enhances marriage prospects, though high search costs from time and resource constraints may limit upward mobility (due to scarce suitable partners), prompt trait exchanges, or lead to 'opt-outs' from marriage market leading a professional life not very different from the male counterpart (Esteve et al., 2020).

**Figure 3: Inverted U Spousal Age Gap**

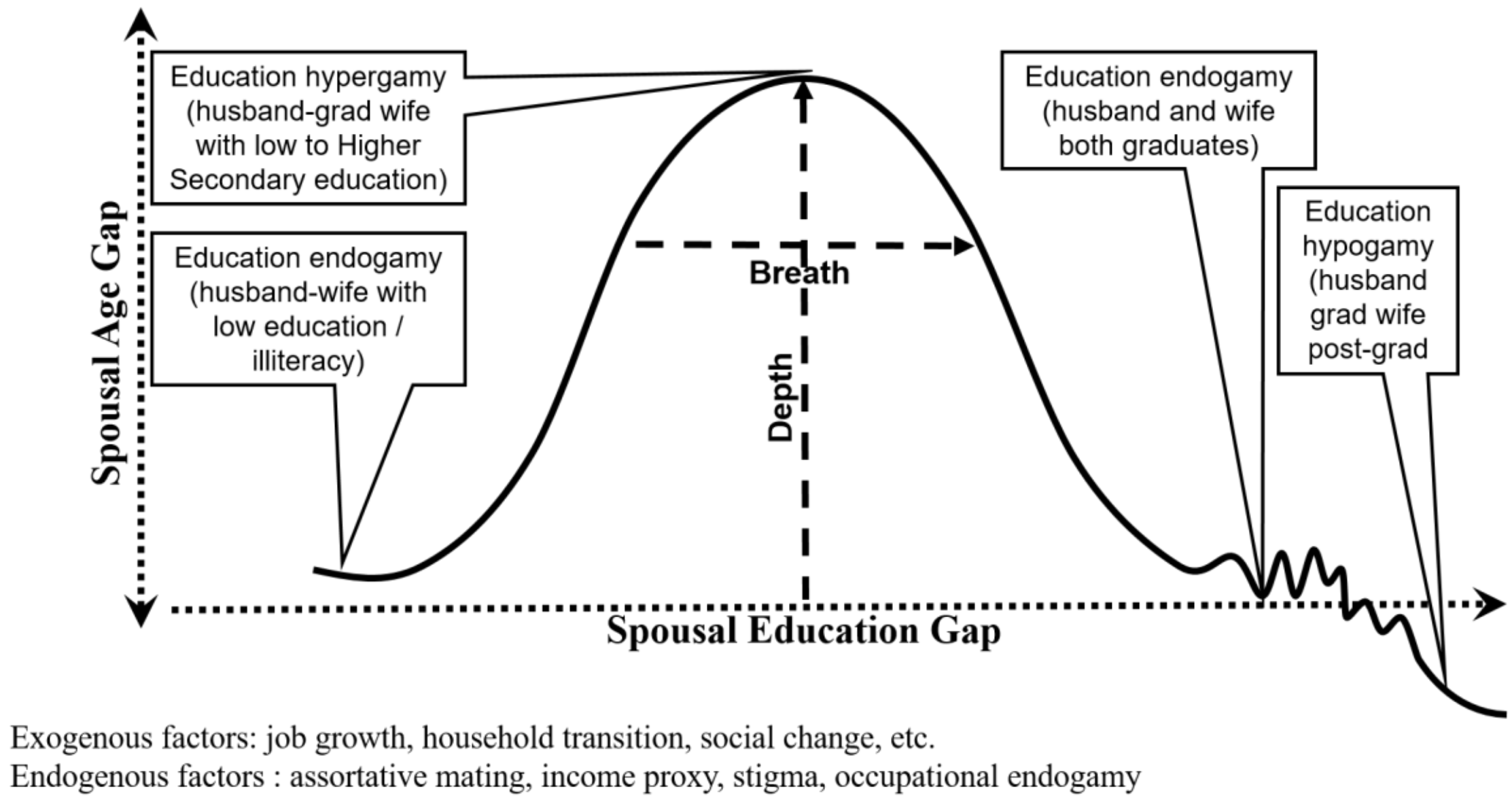

Our inverted-U hypothesis for spousal age gaps by educational differences thus outlines a transitional mechanism akin to the Kuznets curve and Goldin's (1995) U-shaped female labor force function. The net outcome hinges on the relative strength of (1) the "tug of war" between negative income effects and positive substitution effects, (2) the evolution and manifestation of gender norms, and (3) changes in labor market stigma. Additionally, the process is shaped by the availability of jobs, demographic structure—such as the proportion of marriageable women—family structure, social group membership, and regional heterogeneity.

Occupational status serves as a potent marker of economic and cultural capital within households. For example, a white-collar husband in a predominantly blue-collar region, prevailing social norms may penalize women's employment—while raising their reservation wages and diminishing their labor force participation. We interpret this as a localized stigma effect. Conversely, dual white-collar occupations enable homogamy as a gender-egalitarian norm, countering negative income effects and stigma. Here, egalitarian norms may allow substitution effects to dominate, potentially diminishing structural gender constructs like spousal age gaps. Thus, depending on the dominance of these competing forces[7], societies may experience two possible outcomes: one can push up age hypergamy, while the other can pull down the spousal age gap.

## 3. Data

The quinquennial rounds of India's Employment and Unemployment Surveys, conducted by the National Sample Survey (NSS), provide detailed household characteristics at national and state levels, offering a rich basis for analyzing the interplay between marriage and labor markets. Although NSS data lack a predefined couple file, we construct one by organizing household data, grouping individuals, and identifying relational statuses such as household head, spouse, or married child.

[7] However, it remains to be determined which aspect of stigma (a testable argument) primarily influences decision-making, a question requiring further investigation beyond the scope of our current analysis.

We calculate spousal age differences from household files, initially ordering male partners first, then implementing a validation procedure by reversing the gender sequence to ensure robustness. To maintain data integrity, we exclude statistical outliers where age differences fall below -10 years or exceed 25 years. This approach captures the substantial majority of the age difference distribution while eliminating implausible cases.

The NSS provides extensive socio-demographic information including social group affiliation, religious background, and household structure. Since direct income data are unavailable, we use monthly per capita consumption expenditure as a reliable proxy for household economic status. To ensure temporal consistency, we incorporate supplementary data from the 50th round (July 1993-June 1994) and standardize all geographic units to the 1993-94 regional classification, maintaining 78 consistent regions throughout our analysis period.

Our analytical framework restricts the spousal age range to -10 to +25 years to focus on demographically meaningful relationships. Community-level variables are aggregated at the regional level to capture local marriage market dynamics while minimizing selection bias.

Apart from socioeconomic and demographic variables, we categorize female education into three broad categories—primary to higher secondary, graduate, and postgraduate—using illiterate as the reference category, while holding husbands' education at the graduate level. The overall change in economic and household structure is estimated using (1) principal employment status to reflect economic transformation, avoiding subsidiary status to minimize reporting bias and definitional inconsistencies, particularly in rural areas; and (2) a methodology adapted from Ruggles (2012), which focuses on the living arrangements of cohabiting couples to assess household transitions. Community-level variables, such as caste and religion, are aggregated at the regional level.

## 4. Indices to represent Transition of Household Structures, Economic, Cultural and Social Norms

To assess the impact of shifts in household structures, gender norms, occupational opportunities, and labor market dynamics on the spousal age gap (SAG), we develop four indices.

We construct a household structure index following the methodology outlined by Ruggles (2012) and adapted by Breton (2019). Households are classified into five categories: (1) disparate setups, (2) nuclear (comprising one married woman with a husband aged 25–45), (3) supplemented nuclear (one married woman with a husband aged 25–45 and a lone parent aged 65+ from either spouse), (4) stem (one married woman with a husband aged 25–45 and another with a husband aged 65+), and (5) extended (more than two married women with the household head aged 65+). Consistent with Breton (2019), we focus on households with patrifiliative affiliation[8]—defined as bonds established through the paternal line—categorizing them into peripheral (transitory) or core (non-transitory) types based on intergenerational membership. Categories 1 (disparate setups) and 3 (supplemented nuclear setups) are considered transitory, as they may evolve into stable, non-transitory arrangements over time. Disparate setups are characterized by high rates of non-marital births, divorce, and short-lived cohabiting relationships, a pattern more common in Western societies but increasingly visible in urban India. In this context, such arrangements are viewed as transitional downward mobility, serving as short-term configurations for couples before shifting to non-transitory structures, often nuclear households. Similarly, supplemented nuclear setups provide economic benefits, notably guaranteed old-age support, yet remain transitory, transitioning to nuclear households following the lone parent's death. The index, a Bartik-type instrument, combines the proportion of transitory households in the base period with changes between quinquennial NSS rounds.

[8] Patrifiliative affiliation' refers to the bond or affiliation established through a paternal line (Oxford English Dictionary). In a patrilocal residential context, household types can be distinguished primarily based on their core

We assume that state-level growth rates of transitory households are independent of region-level transitions, which are driven by demographic shifts, settlement patterns, age structures, and gender/caste/elderly compositions[9]. To minimize selectivity bias, we exclude the region-specific proportion of transitory households from state-level estimates, ensuring the index captures exogenous regional variations.

A parallel Bartik-type index measures structural change in white-collar job growth, integrating a stock variable (proportion of white-collar jobs in the base period) with a flow variable (change between quinquennial rounds). This approach, employed by Blanchard and Katz (1992), Bound and Holzer (2000), Autor and Duggan (2003), and Datta Gupta et al. (2007), addresses selectivity concerns in labor market analyses.

We also develop two proxy variables to reflect influences on women's autonomy. The first, a stigma index, captures barriers to female labor force participation, drawing on Goldin's (1995) observation that "no educated, high-income man would permit his wife to work in manufacturing." This index is constructed at the NSS region level, identifying households where husbands hold white-collar jobs in areas where 75% of employment is blue-collar, limiting wives' opportunities. The second, a gender-egalitarian norm index, quantifies occupational homogamy—the percentage of couples both in white-collar roles—reflecting prestige, scholastic achievement, economic stability, and social standing. In patriarchal settings, such homogamy enables women to gain recognition beyond traditional roles, promoting gender equality.

Cultural influences on hypergamy extend beyond individual community membership to the dominance of certain groups. Our caste index, a Bartik-type measure, combines cross-regional differences in the non-scheduled caste/non-scheduled tribe population with state-level changes between rounds, capturing exogenous cultural forces.

Spatial heterogeneity in India's female deficit, noted by Miller (1981), Dyson and Moore (1983), Sopher (1980), Agnihotri (2000), and Siddhanta et al. (2003), shapes SAG variations. We account for this with a bride deficit index, a Bartik-type instrument multiplying the change

---

membership, which comprises married couples related by patrifiliation (Breton 2019). Households that lack a patrifiliative intergenerational core are classified as a peripheral category.

[9] Demographic change in India is more localized than structural. For instance, the pattern of fertility decline or the masculinization of the child sex ratio in India displays a higher degree of within-state diversity than between-state differences (Guilmoto & Vaguet 2000, Nandy & Siddhanta 2014, etc.). This squarely aligns with our perspective that change in household structure in India are more localized or regional (e.g., agro-climatic regions, districts, tehsils, etc.) than confined by state boundaries.

in the marriageable-age female cohort by its base-period proportion. This approach reflects localized demographic shifts, consistent with within-state diversity in fertility decline and child sex ratio masculinization

## 5. Empirical Model and Results

### Modelling spousal age gap

Building on the theoretical framework outlined in Section 2, we investigate four primary mechanisms influencing the spousal age gap (SAG): (1) income effects, proxied by household consumption expenditure; (2) education matching, represented by graduate husbands paired with wives across educational levels (primary to higher secondary, graduate, or postgraduate); (3) transitory forces, captured by changes in white-collar job growth and household structures; and (4) social norms, including stigma (which widens SAG by restricting female labor force participation) and occupational homogamy (which mitigates patriarchal constraints). We control for religion, caste, and region as potential confounders.

The empirical model employs an OLS regression, specifying SAG as a function of household consumption expenditure (categorized into 12 monthly per capita expenditure [MPCE] fractile classes), education assortative mating (dummy variables for wife's education levels), stigma (husbands in white-collar jobs in regions where 75% of employment is blue-collar), the household structure transition index (proportionate change in transitory households), the occupational homogamy index, the bride deficit index (change in the marriageable-age female cohort weighted by its base-period proportion), and network- or group-affiliation factors. Regional fixed effects account for India's five major regions—North, South, East, West, and Northeast—with Central as the reference category. To address the persistent influence of arranged marriages on age hypergamy, we include the regional average partner age gap from the previous decade as a control to estimate the precise impact of each explanatory variable, particularly education matching variables. Robust standard errors, clustered at the regional level, address potential heteroskedasticity arising from regional variations. However, endogeneity—such as reverse causality where SAG might influence education matching—remains a limitation, mitigated through sensitivity analyses.

### Descriptive Statistics

Table 1 presents comprehensive descriptive statistics illustrating the demographic and socioeconomic transitions across rural and urban India between the 61st (2004-05) and 68th (2011-12) NSS rounds. The average spousal age gap declined modestly from 4.97 to 4.85 years in rural areas and from 5.28 to 5.06 years in urban areas, signalling a gradual shift in marriage customs, though urban India maintains consistently larger age gaps.

Educational Assortative Mating Patterns: Educational dynamics reveal significant changes in marital matching. Educational hypergamy (graduate husbands with primary to high-secondary educated wives) increased from 4.8% to 6.7% in rural areas and from 10.3% to 12.3% in urban areas. Educational homogamy (both spouses graduates) rose from 1.0% to 1.6% rurally and from 5.4% to 6.8% in urban sectors. Most notably, educational hypogamy (graduate husbands with postgraduate wives) expanded dramatically from 0.1% to 0.6% in rural areas and from 1.1% to 3.4% in urban areas, indicating growing female educational advancement.

## Table 1: Variable definitions and means

| | Rural Sector | | Urban Sector | |
|---|---|---|---|---|
| | 61st round (2004-2005) | 68th round (2011-2012) | 61st round (2004-2005) | 68th round (2011-2012) |
| *Outcome variable*: Spousal Age Gap | 4.969<br>(3.060) | 4.847<br>(2.993) | 5.276<br>(3.227) | 5.064<br>(3.166) |
| *Key Explanatory Variables* | | | | |
| Husband grad & wife primary to high-secondary education | 0.048<br>(0.214) | 0.067<br>(0.250) | 0.103<br>(0.304) | 0.123<br>(0.328) |
| Husband and wife graduate | 0.010<br>(0.100) | 0.016<br>(0.124) | 0.054<br>(0.226) | 0.068<br>(0.252) |
| Husband grad & wife postgrad | 0.001<br>(0.033) | 0.006<br>(0.075) | 0.011<br>(0.105 | 0.034<br>0.181) |
| Stigma: | | | | |
| Husband in white-collar job in blue-collar region | 0.100<br>(0.300) | 0.146<br>(0.353) | 0.187<br>(0.390) | 0.269<br>(0.444) |
| Structure: | | | | |
| Index of white-collar job growth in region | -0.046<br>(0.046) | 0.004<br>(0.069) | -0.043<br>(0.047) | 0.006<br>(0.068) |
| Index of Household transition | -0.252<br>(0.090) | -0.009<br>(0.017) | -0.248<br>(0.094) | -0.009<br>(0.019) |
| Index of female deficit in marriageable age cohort (bride deficit) | -0.012<br>(0.025) | 0.003<br>(0.024) | -0.014<br>(0.025) | 0.001<br>(0.027) |
| Index of growth of nonscst population: | -0.046<br>(0.134) | 0.011<br>(0.073) | -0.037<br>(0.112) | 0.023<br>(0.106) |
| Occupation homogamy | 0.037<br>(0.189) | 0.018<br>(0.133) | 0.048<br>(0.215) | 0.042<br>(0.201) |
| *Control* | | | | |
| average partner age gap at the regional level from the previous decade | 5.394<br>(1.279) | 5.082<br>(1.263) | 5.44<br>(1.297) | 5.079<br>(1.276) |
| household head's age | 37.104<br>(12.492) | 38.262<br>(12.379) | 37.388<br>(11.824) | 38.741<br>(11.984) |
| household head's agesquared | 1532.772<br>(1038.179) | 1617.172<br>(1049.939) | 1537.655<br>(991.878) | 1644.495<br>(1030.801) |
| MPCE fractile class 1 | 0.020<br>(0.140) | 0.034<br>(0.182) | 0.049<br>(0.216) | 0.053<br>(0.224) |
| MPCE fractile class 2 | 0.023<br>(0.150) | 0.030<br>(0.172) | 0.053<br>(0.224) | 0.061<br>(0.239) |
| MPCE fractile class 3 | 0.056<br>(0.230) | 0.066<br>(0.248) | 0.107<br>(0.309) | 0.113<br>(0.317) |
| MPCE fractile class 4 | 0.073<br>(0.260) | 0.079<br>(0.269) | 0.127<br>(0.332) | 0.098<br>(0.297) |
| MPCE fractile class 5 | 0.080<br>(0.272) | 0.083<br>(0.276) | 0.102<br>(0.303) | 0.090<br>(0.287) |
| MPCE fractile class 6 | 0.085<br>(0.279) | 0.087<br>(0.281) | 0.097<br>(0.296) | 0.092<br>(0.289) |
| MPCE fractile class 7 | 0.102<br>(0.302) | 0.099<br>(0.299) | 0.087<br>(0.282) | 0.094<br>(0.292) |
| MPCE fractile class 8 | 0.118 | 0.112 | 0.078 | 0.090 |
| | (0.322) | (0.316) | (0.269) | (0.286) |
| MPCE fractile class 9 | 0.140 | 0.120 | 0.090 | 0.100 |
| | (0.347) | (0.325) | (0.287) | (0.300) |
| MPCE fractile class 10 | 0.146 | 0.136 | 0.104 | 0.112 |
| | (0.353) | (0.343) | (0.306) | (0.315) |
| MPCE fractile class 11 | 0.085 | 0.073 | 0.059 | 0.059 |
| | (0.278) | (0.261) | (0.236) | (0.235) |
| MPCE fractile class 12 | 0.074 | 0.080 | 0.046 | 0.037 |
| | (0.261) | (0.271) | (0.210) | (0.190) |
| Socioreligious group is Islam | 0.102 | 0.118 | 0.141 | 0.155 |
| | (0.302) | (0.322) | (0.348) | (0.362) |
| Socioreligious group is Christian | 0.067 | 0.068 | 0.066 | 0.062 |
| | (0.251) | (0.252) | (0.248) | (0.242) |
| Socioreligious group is Sikh | 0.033 | 0.025 | 0.02 | 0.021 |
| | (0.180) | (0.156) | (0.139) | (0.142) |
| Socioreligious group is Buddhist | 0.012 | 0.011 | 0.008 | 0.007 |
| | (0.107) | (0.105) | (0.089) | (0.083) |
| Region is North | 0.157 | 0.162 | 0.160 | 0.178 |
| | (0.363) | (0.368) | (0.367) | (0.382) |
| Region is East | 0.115 | 0.114 | 0.097 | 0.094 |
| | (0.320) | (0.318) | (0.296) | (0.292) |
| Region is West | 0.103 | 0.110 | 0.168 | 0.152 |
| | (0.304) | (0.313) | (0.373) | (0.359) |
| Region is South | 0.193 | 0.193 | 0.255 | 0.246 |
| | (0.394) | (0.395) | (0.255) | (0.431) |
| Region is Northeast | 0.146 | 0.144 | 0.106 | 0.117 |
| | (0.353) | (0.351) | (0.308) | (0.321) |
| Observations | 97121 | 69482 | 45151 | 41206 |

Note: standard deviations in parentheses. Educational variables are dummy indicators (1 = condition met, 0 = otherwise). Indices are standardized Bartik-type measures (mean 0, variance adjusted by round).

Labor Market and Social Dynamics: The phenomenon of stigma—defined here as a husband in a white-collar job residing in a blue-collar region—also showed a rising trend from 2004-2005 to 2011-2012, with a higher increase in the urban sector. Likewise, the mean values for job growth index, household transition index, and the index of bride deficit generally increased in both sectors over this period. The mean of growth for the non-SC-non-ST population also increased, albeit marginally. The distribution across monthly per capita consumption expenditure (MPCE) levels indicate a slight increase at higher levels for both urban and rural sectors, suggesting improvements in consumption levels over time. The proportion of Muslims increased slightly in both sectors while the proportions of the other minority religious groups (Christians, Sikhs, and Buddhists) remained relatively stable.

The distribution of households across regions-maintained consistency over time in both sectors. The South is more urbanised, while the household share in North and North-East saw minor increases in the urban sectors between 2004-05 and 2011-12.

**Regression results:**

We estimate a simple OLS regression. The only covariates are education matching variables which are graduate husband and wife with different educational levels, which include primary, middle, secondary, high secondary, graduate, and postgraduate. When no controls are added, a basic non-linear decreasing trend emerges in the rural sector, with the highest significant coefficient when the husband is a graduate and the wife has primary to high-secondary education (pri+mid+sec+hs). In the base model, when both the husband and wife are graduates, the coefficient is statistically insignificant in both rounds. Graduate husband with postgraduate wife has a significant negative effect implying that educational hypogamy is inversely related to the spousal age gap. potentially indicating complementarity of traits in cases of negative assortative mating (e.g., wife's education vs. husband's income), which is a plausible assumption given that individuals often balance unequal traits through exchange. For example, relatively low-educated men with class privilege can exchange their class to marry women from economically poor but educationally rich (education or caste) backgrounds. The outcome of such matches is negative assortative mating, which can arise as people exchange one advantage with another (Davis 1941; Merton 1941).

**Table 2a**: OLS model of the effect of education matching, stigma, structure, household transition, female deficit and MPCE classes on spousal age gap: rural sector

| **Rural** | **61st round (2004-2005)** | | | | **68th round (2011-2012)** | | | |
|---|---|---|---|---|---|---|---|---|
| | no control | + stigma & structure | + prestige | + all controls | no control | + stigma & structure | + prestige | + all controls |
| Husband grad & wife primary to high-secondary education | 0.410*** (0.0456) | 0.265*** (0.0474) | 0.249*** (0.0474) | 0.397*** (0.0432) | 0.403*** (0.0460) | 0.245*** (0.0484) | 0.248*** (0.0485) | 0.419*** (0.0437) |
| Husband and wife graduate | -0.0582 (0.0945) | -0.272*** (0.0970) | -0.229** (0.0970) | -0.219** (0.0957) | -0.0710 (0.0959) | -0.240** (0.0971) | -0.259*** (0.0988) | -0.232** (0.0917) |
| Husband grad & wife postgrad | -0.441* (0.232) | -0.881*** (0.233) | -0.834*** (0.232) | -0.593*** (0.222) | -0.454*** (0.150) | -0.718*** (0.152) | -0.753*** (0.155) | -0.528*** (0.151) |
| Stigma: | | | | | | | | |
| Husband in white-collar job in blue-collar region | | 0.146*** (0.0346) | 0.256*** (0.0394) | 0.180*** (0.0363) | | 0.104*** (0.0347) | 0.0926*** (0.0351) | 0.0447 (0.0327) |
| Structure: | | | | | | | | |
| Index of white-collar job growth in region | | -6.771*** (0.239) | -6.674*** (0.239) | 0.323 (0.264) | | -1.028*** (0.206) | -1.044*** (0.206) | 0.211 (0.214) |
| Household transition | | 1.299*** (0.113) | 1.255*** (0.113) | 2.526*** (0.167) | | -2.648*** (0.668) | -2.639*** (0.668) | -1.391** (0.644) |
| Female deficit in marriageable age cohort (bride deficit) | | -9.724*** (0.441) | -9.547*** (0.442) | -3.566*** (0.472) | | 11.83*** (0.505) | 11.85*** (0.505) | 1.063** (0.532) |
| MPCE fractile class 2 | | 0.00675 (0.0823) | 0.00749 (0.0823) | 0.0428 (0.0746) | | 0.110 (0.0790) | 0.110 (0.0790) | 0.0511 (0.0730) |
| MPCE fractile class 3 | | 0.00144 (0.0729) | 0.00243 (0.0729) | 0.133** (0.0662) | | 0.251*** (0.0666) | 0.251*** (0.0666) | 0.0858 (0.0612) |
| MPCE fractile class 4 | | 0.130* (0.0703) | 0.132* (0.0703) | 0.195*** (0.0638) | | 0.396*** (0.0660) | 0.396*** (0.0660) | 0.0874 (0.0611) |
| MPCE fractile class 5 | | 0.171** (0.0701) | 0.173** (0.0701) | 0.189*** (0.0639) | | 0.414*** (0.0657) | 0.414*** (0.0657) | 0.0601 (0.0610) |
| MPCE fractile class 6 | | 0.268*** (0.0698) | 0.270*** (0.0698) | 0.266*** (0.0637) | | 0.583*** (0.0662) | 0.583*** (0.0662) | 0.233*** (0.0614) |
| MPCE fractile class 7 | | 0.277*** (0.0684) | 0.279*** (0.0684) | 0.252*** (0.0627) | | 0.583*** (0.0648) | 0.583*** (0.0648) | 0.233*** (0.0605) |
| MPCE fractile class 8 | | 0.354*** (0.0686) | 0.356*** (0.0686) | 0.320*** (0.0631) | | 0.511*** (0.0632) | 0.510*** (0.0632) | 0.184*** (0.0593) |
| MPCE fractile class 9 | | 0.471*** (0.0676) | 0.473*** (0.0676) | 0.356*** (0.0623) | | 0.591*** (0.0632) | 0.590*** (0.0632) | 0.261*** (0.0592) |
| MPCE fractile class 10 | | 0.481*** (0.0679) | 0.483*** (0.0679) | 0.396*** (0.0627) | | 0.709*** (0.0629) | 0.708*** (0.0629) | 0.332*** (0.0596) |
| MPCE fractile class 11 | | 0.433*** (0.0722) | 0.440*** (0.0722) | 0.386*** (0.0669) | | 0.766*** (0.0709) | 0.764*** (0.0710) | 0.397*** (0.0671) |
| MPCE fractile class 12 | | 0.663*** (0.0748) | 0.673*** (0.0748) | 0.506*** (0.0698) | | 0.902*** (0.0708) | 0.899*** (0.0709) | 0.463*** (0.0676) |
| Prestige: | | | | | | | | |
| Occupation homogamy | | | -0.360***<br>(0.0599) | -0.464***<br>(0.0579) | | | 0.143<br>(0.103) | -0.193**<br>(0.0980) |
| Constant | 4.951***<br>(0.0101) | 4.547***<br>(0.0679) | 4.542***<br>(0.0678) | -1.619***<br>(0.111) | 4.824***<br>(0.0119) | 4.221***<br>(0.0541) | 4.222***<br>(0.0541) | -1.331***<br>(0.119) |
| R-squared | 0.001 | 0.019 | 0.020 | 0.195 | 0.001 | 0.015 | 0.015 | 0.183 |
| Observations | 97,121 | 95,623 | 95,623 | 95,623 | 69,482 | 69,482 | 69,482 | 69,482 |

*Note***:** Other controls include: intercept, age of the household head, age-square, caste, religion and region indicators. Omitted religion category is Hindu, omitted caste category is General, omitted region category is Central. ***1% level, **5% level, *10% level significance. Abbreviation**:** OLS, Ordinary Least Square; MPCE, Monthly Per Capita Consumption Expenditure.

**Table 2b**: OLS model of the effect of education matching, stigma, structure, household transition, female deficit and MPCE classes on spousal age gap: urban sector

| **Urban** | **61st round (2004-2005)** | | | | **68th round (2011-2012)** | | | |
|---|---|---|---|---|---|---|---|---|
| | no control | + stigma & structure | + Prestige | + all controls | no control | + stigma & structure | + Prestige | + all controls |
| Husband grad & wife primary to high-secondary education | 0.390***<br>(0.0497) | 0.309***<br>(0.0526) | 0.286***<br>(0.0526) | 0.332***<br>(0.0501) | 0.328***<br>(0.0481) | 0.228***<br>(0.0506) | 0.210***<br>(0.0506) | 0.294***<br>(0.0473) |
| Husband and wife graduate | -0.491***<br>(0.0598) | -0.677***<br>(0.0645) | -0.620***<br>(0.0646) | -0.466***<br>(0.0620) | -0.299***<br>(0.0570) | -0.466***<br>(0.0611) | -0.410***<br>(0.0613) | -0.333***<br>(0.0578) |
| Husband grad & wife postgrad | -0.837***<br>(0.134) | -1.045***<br>(0.137) | -0.963***<br>(0.138) | -0.548***<br>(0.135) | -0.855***<br>(0.0764) | -1.077***<br>(0.0813) | -0.939***<br>(0.0831) | -0.556***<br>(0.0801) |
| Stigma: | | | | | | | | |
| Husband in white-collar job in blue-collar region | | -0.0617<br>(0.0402) | 0.0227<br>(0.0414) | 0.146***<br>(0.0394) | | -0.0735**<br>(0.0367) | -0.0228<br>(0.0370) | 0.0117<br>(0.0353) |
| Structure: | | | | | | | | |
| Index of white-collar job growth in region | | -2.401***<br>(0.355) | -2.299***<br>(0.355) | -1.088***<br>(0.369) | | 0.525*<br>(0.288) | 0.551*<br>(0.288) | -0.180<br>(0.270) |
| Household transition | | 3.478***<br>(0.188) | 3.406***<br>(0.188) | 2.568***<br>(0.304) | | -0.0885<br>(0.801) | -0.139<br>(0.800) | -0.166<br>(0.858) |
| Female deficit in marriageable age cohort (bride deficit) | | -5.904***<br>(0.677) | -5.733***<br>(0.677) | -3.632***<br>(0.697) | | 7.962***<br>(0.581) | 7.902***<br>(0.579) | -0.297<br>(0.662) |
| MPCE fractile class 2 | | 0.0923<br>(0.0869) | 0.0946<br>(0.0869) | 0.0465<br>(0.0799) | | 0.206**<br>(0.0852) | 0.206**<br>(0.0852) | 0.148*<br>(0.0801) |
| MPCE fractile class 3 | | 0.263***<br>(0.0761) | 0.265***<br>(0.0761) | 0.0746<br>(0.0709) | | 0.205***<br>(0.0771) | 0.206***<br>(0.0771) | 0.158**<br>(0.0726) |
| MPCE fractile class 4 | | 0.345***<br>(0.0756) | 0.343***<br>(0.0757) | 0.0775<br>(0.0710) | | 0.243***<br>(0.0792) | 0.245***<br>(0.0792) | 0.162**<br>(0.0748) |
| MPCE fractile class 5 | | 0.350***<br>(0.0781) | 0.350***<br>(0.0781) | 0.133*<br>(0.0734) | | 0.298***<br>(0.0804) | 0.300***<br>(0.0804) | 0.238***<br>(0.0763) |
| MPCE fractile class 6 | | 0.447***<br>(0.0814) | 0.449***<br>(0.0814) | 0.265***<br>(0.0769) | | 0.439***<br>(0.0825) | 0.442***<br>(0.0825) | 0.390***<br>(0.0780) |
| MPCE fractile class 7 | | 0.583***<br>(0.0827) | 0.587***<br>(0.0828) | 0.383***<br>(0.0783) | | 0.347***<br>(0.0802) | 0.349***<br>(0.0801) | 0.305***<br>(0.0769) |
| MPCE fractile class 8 | | 0.531***<br>(0.0839) | 0.534***<br>(0.0840) | 0.324***<br>(0.0798) | | 0.393***<br>(0.0822) | 0.403***<br>(0.0822) | 0.313***<br>(0.0785) |
| MPCE fractile class 9 | | 0.557***<br>(0.0821) | 0.563***<br>(0.0821) | 0.295***<br>(0.0780) | | 0.435***<br>(0.0818) | 0.452***<br>(0.0817) | 0.352***<br>(0.0787) |
| MPCE fractile class 10 | | 0.654***<br>(0.0815) | 0.675***<br>(0.0815) | 0.311***<br>(0.0787) | | 0.655***<br>(0.0805) | 0.673***<br>(0.0805) | 0.395***<br>(0.0781) |
| MPCE fractile class 11 | | 0.513***<br>(0.0915) | 0.555***<br>(0.0916) | 0.142<br>(0.0893) | | 0.767***<br>(0.0930) | 0.792***<br>(0.0930) | 0.406***<br>(0.0906) |
| MPCE fractile class 12 | | 0.507***<br>(0.0975) | 0.570***<br>(0.0977) | 0.0805<br>(0.0959) | | 0.723***<br>(0.109) | 0.770***<br>(0.109) | 0.297***<br>(0.106) |
| Prestige: | | | | | | | | |
| Occupation homogamy | | | -0.581***<br>(0.0721) | -0.718***<br>(0.0702) | | | -0.632***<br>(0.0784) | -0.628***<br>(0.0755) |
| Constant | 5.272***<br>(0.0168) | 5.588***<br>(0.0808) | 5.579***<br>(0.0808) | 0.374**<br>(0.179) | 5.073***<br>(0.0179) | 4.727***<br>(0.0632) | 4.724***<br>(0.0632) | -0.768***<br>(0.174) |
| R-squared | 0.003 | 0.019 | 0.020 | 0.139 | 0.004 | 0.013 | 0.014 | 0.150 |
| Observations | 45,151 | 44,705 | 44,705 | 44,705 | 41,206 | 41,206 | 41,206 | 41,206 |

*Note***:** Other controls include: intercept, age of the household head, age-square, caste, religion and region indicators. Omitted religion category is Hindu, omitted caste category is General, omitted region category is Central. ***1% level, **5% level, *10% level significance.

Abbreviation: OLS, Ordinary Least Square; MPCE, Monthly Per Capita Consumption Expenditure.

Upon adding social norms, occupational opportunities, index of household structure transition and bride deficit, the basic relationship between educational matching and age hypergamy remains unchanged, indicating uncorrelatedness between educational matching and these factors. It is important to note that both stigma and index of gender composition have the expected positive and negative sign while the sign of the index of structural change, though negative during 2004-05, became positive in 2011-12. Furthermore, the index of household transition was positive and significant during 2004-05 *but insignificant in the urban sector during 2011-12*. The relationship between income and age hypergamy (after adding all major covariates) is positive significant in both the rounds.

Between the two successive rounds, the locus of educational matching and age hypergamy shifts rightward, and during 2011-12, the coefficient of graduate husband and wife is, though insignificant in no-control model, but significant in other model specifications.

Next, we incorporate the index of occupational homogamy (measure of prestige) to judge the relative impact of gender equality in labor market outcome on the overall relationship between husband-wife educational matching and their age difference. Once again, the basic relationship holds. The coefficients for graduate husband with post-graduate wife is significant and negative in both rounds. The stigma coefficient drops only marginally after incorporating occupational homogamy (our measure of prestige) in the regression model.

The synthesis (all control) model for the rural sector shows that education hypergamy (graduate husband with primary to high secondary educated wife) is more than 40 percent more likely to expand age hypergamy compared to graduate husband with illiterate wife. However, educational hypogamy has strong inverse association with age hypergamy in both rounds of the rural sector. A notable reduction in the stigma effect is observed between the 61$^{st}$ and 68$^{th}$ rounds. More importantly the index of household structure is significant and negative in 68$^{th}$ round implying that the growth of transitional households as driver of cultural change can reduce age hypergamy. The Index of bride deficit measured as change in availability of females in marriageable age cohort is though negative and significant during 2004-05 (squeeze effect), but becomes positive and significant in 2011-12 (mobility or income effect), which may imply that the adaptability of societal practices and bargaining power of females in marriage market in response to bride squeeze can change with time.

The impact of the marital age gap of 10 years lag is significant, with a coefficient value of 0.875 in the 61$^{st}$ round and 0.865 in the 68$^{th}$ round, substantiating the intergenerational nature of Indian marriage. Caste and religion show expected signs and regional differences are significant; eastern and southern India are pushing up the age gap in both rounds, while northern and western India are showing an inverse association with age hypergamy in the 68$^{th}$ round (vide appendix 4.1(a)).

Even after controlling for all factors, the income effect, (the coefficients of income fractile classes) is significant in both the urban and rural sectors. The coefficient plots of fractile classes (appendix 4) indicate a rising adverse (positive) income effect on spousal age gap in the rural sector, while in the urban sector, the impact is non-linear, rising up to the sixth fractile class, followed by a stalling pattern up to the ninth fractile class, and then a decline in the coefficient value (the coefficient of the twelfth fractile class, though positive and significant, is lower than the corresponding coefficient at the seventh fractile class), indicating a sign of gender-supportive change in the income effect at high level of prosperity in the urban sector.

**Figure 4**: Coefficient plots of rural and urban sectors (61st and 68th rounds): ordinary least squares (OLS) estimates

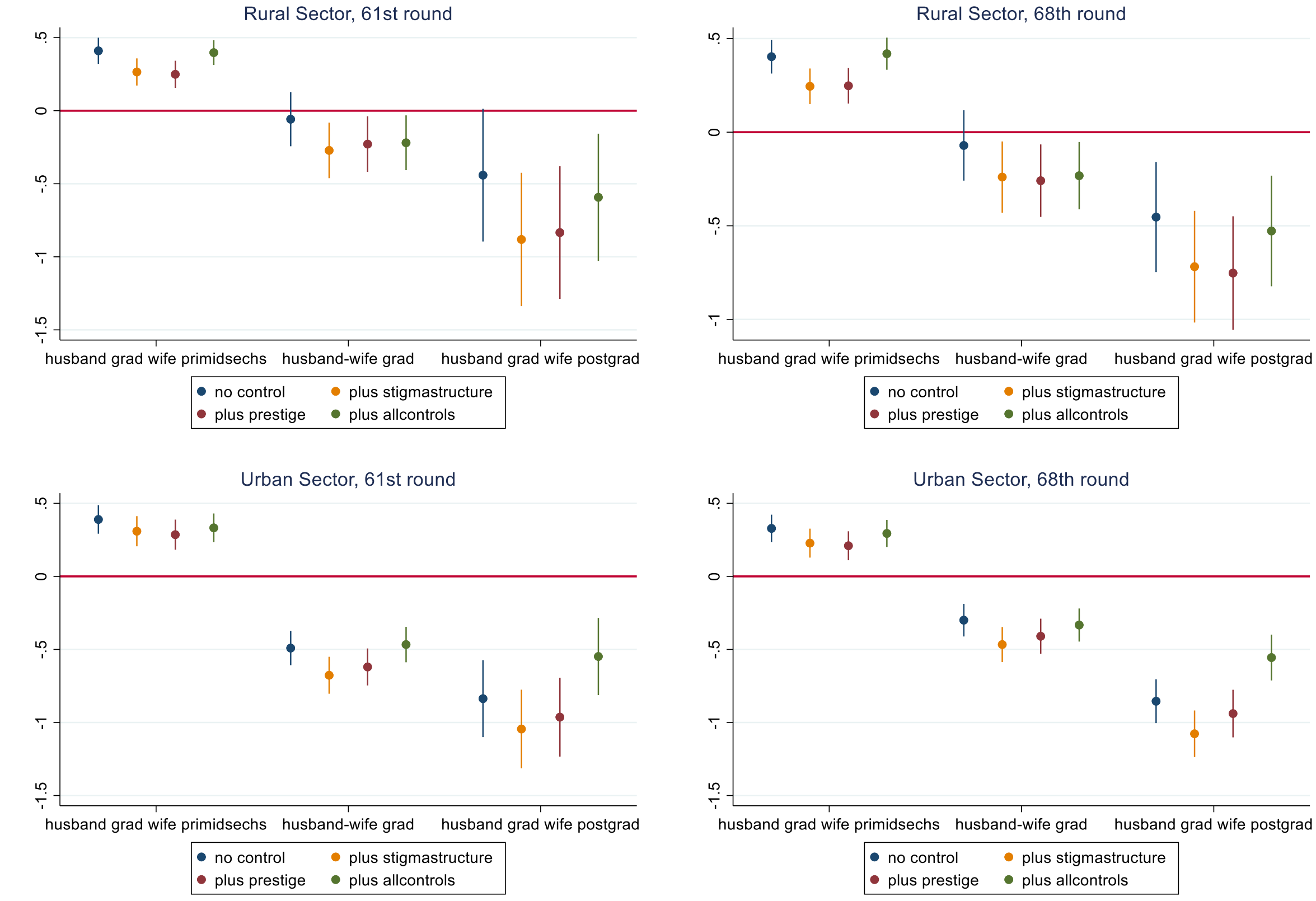

Similar to rural sector, urban sample from 61st and 68th round is also showing an inverse association with husband-grad and wife-post-grad and spousal age gap. The pattern of the relationship is more pronounced in 68th round with a sharper fall in coefficient if we transit from education hypergamy to homogamy, and then a much sharper fall when we transit from homogamy to hypogamy. This basic relationship holds both in 61st and 68th rounds, if we add stigma, structure, household structure and index of bride deficit. After incorporating the index of occupation homogamy there is some improvement in the goodness of fit of the model and the coefficient is negative significant in both the rounds. Nevertheless, incorporation of one more significant factor does not hamper the basic relationship between education matching and husband-wife age gap. In the final synthesis model, the basic relationship remains. Transiting from education hypergamy to education hypogamy can reduce age hypergamy significantly even though the impact of per capita consumption expenditure is positive – significant upto 6 or 7th fractile class along with negative significant impact of structural change (job index). While the index of missing brides and household structure are significant during 2004-05, they become insignificant in 2011-12. Occupational homogamy was significant-negative in both rounds. The lag effect of regional spousal age gap was more prominent in 68th round with the coefficient of 0.82 as compared to the corresponding 0.68 in 61st round.

The findings of our study confirm that assortative mating and an income effect, in conjunction with societal stigma and respect (prestige), collectively shape the contemporary trends of age hypergamy in Indian society.

**Role of Unobservables**

However, how strong does selection on unobservables have to be relative to selection on observables in order to nullify the treatment effect?

We employ the method originally introduced by Altonji et al. (2005) and further developed by Oster (2017, 2019), Diegert et al. (2022) and Masten & Poirier (2022) to assess the importance of selection on unobservables, with the key assumption that selection in unobservables is proportional to the selection in observables. This method provides break-down point, i.e., the maximum extent to which an assumption can be violated without causing misleading or erroneous results. The method is operationalized by assessing the degree to which the treatment effects (in this case, the coefficients of education matching) change when controls are added, scaled by R-squared. The crucial step in sensitivity analysis is determining the maximum R-squared value, which is the highest possible R-squared that could be achieved by including unobservable variables in the model. In line with Oster's (2017, 2019) proposition, we accept an R-max value (derived from the most comprehensive model utilizing all controls) equal to 1.3 times the R-squared.

**Table 3**: Breakdown points (in percent)

| | **Rural Sector** | | **Urban Sector** | |
|---|---|---|---|---|
| | $61^{st}$ round | $68^{th}$ round | $61^{st}$ round | $68^{th}$ round |
| Husband Graduate Wife Pri+mid+sec+hs | 100 | 84.4 | 18.5 | 42.4 |
| Husband and wife both Graduates | 26.2 | 64.4 | 53.7 | 39 |
| Husband Graduate Wife Post-graduate | 56.3 | 100 | 19 | 19.7 |

The table (3) present the sensitivity analysis using Oster's (2019) bounds (based on the difference in the treatment effect between a no-control and full-control regression). The breakdown points indicate the total variance in the outcome variable that needs to be explained by omitted variables to fully account for the observed treatment effect. Our estimated breakdown points range from 18.5% (in the urban sector of the $61^{st}$ round) to 100% (in the rural sector of the $61^{st}$ round) when the husband is a graduate and the wife is educated from primary to post-graduate level. In the urban sector, selection bias is more pronounced when the husband is a graduate and the wife is a post-graduate. This dynamic also holds true for education endogamy (with a decline in breakdown point), and education hypergamy, with a rising delta parameter, albeit still on the lower end. Conversely, in the rural sector where the unobservables would have to be much larger than the observables (i.e. strong treatment effect) in order to negate the observed treatment effect (with high and rising delta parameter).

The sensitivity of treatment effects to unobservables is summarized in the Table 3 and Appendix graphs 6a and 6b. In the rural sector (61st round, 2004–05), the effect of a graduate husband with a primary to high-secondary-educated wife on spousal age gap is robust, with a consistently positive beta lower bound across all delta values and no sign-change breakdown point. Stable beta bounds as delta increases reinforce this robustness. The 68th round (2011–12) shows similar consistency, with a breakdown point of 84.4%, indicating low sensitivity to unobservables.

For education homogamy in the rural sector, the treatment effect on age gap is less robust. The breakdown point rises from 26.6% (61st round) to 64.4% (68th round), but wider beta bounds at higher delta values suggest sensitivity to unobserved confounders, warranting further exploration of omitted variables in education endogamy's impact on age-hypergamy.

Educational hypogamy (graduate husband, post-graduate wife) in the rural sector (68th round) exhibits a robust negative effect on age-hypergamy, with a 100% breakdown point and no sign change across delta values. In the 61st round, the negative lower bound persists, but the upper bound flips sign at a 60% breakdown point, signaling potential omitted variable bias.

In the urban sector, both rounds show high sensitivity to unobservables. Sensitivity plots reveal widening beta bounds as delta increases, with low breakdown points and sign-flipping for education homogamy (delta > 0.6 in 61st round, > 0.5 in 68th round) and hypogamy. For hypogamy, beta coefficients change sign, and confidence intervals widen at higher delta values, with low breakdown points confirming substantial omitted variable bias. Higher breakdown points in the 61st round relative to the 68th indicate increasing selectivity bias over time.

The sensitivity analysis performed in this section emphasises:

1. The selection bias is more pronounced in the urban sector, particularly in cases of education hypogamy and homogamy. Despite increase in the delta parameter in the 68$^{th}$ round (18.5 in 61$^{st}$ round to 42.4 in 68$^{th}$ round) selection bias is substantial even in the case of education hypergamy.
2. In the rural sector, selection bias is a matter of concern only in the case of education homogamy.

### Instrument variable

To address selection bias in educational matching, we extend the sensitivity analysis of unobservables with an instrumental variable (IV) approach for rural and urban sectors. In the rural sector (61st round, 2004–05), educational homogamy reflects ascribed characteristics, with prevalence rising from 0.010 to 0.016 (2004–05 to 2011–12), indicating traditional gender exchange. For graduate-level homogamy, we use regional-level[10] educational matching multiplied by female white-collar workforce participation across 78 NSS regions as the IV. Given the cultural and agro-climatic homogeneity of NSS regions and limited inter-regional marriage in rural India before 2000 (Rosenzweig & Stark, 1989; Agnihotri, 1996, 2000; Datta Gupta et al., 2020), this IV is plausibly exogenous. The IV regression yields a high Cragg-Donald F-statistic and an insignificant beta coefficient for the treatment variable (graduate-graduate), indicating robust identification.

**Table 4a**: 2SLS – IV model of the effect of education matching, stigma, structure, household transition, female deficit and MPCE classes on spousal age gap: rural sector

| | **61$^{st}$ round** | **68$^{th}$ round** |
|---|---|---|
| | all controls | all controls |
| Husband and wife graduate | 0.830<br>(5.483) | 0.745<br>(2.344) |
| Husband grad & wife postgrad | -0.565<br>(0.421) | -0.315<br>(0.265) |
| Husband graduate & wife primary/middle/secondary/higher secondary level of education | 0.549***<br>(0.190) | 0.546***<br>(0.110) |
| Stigma: | | |
| Husband in white-collar job in blue-collar region | 0.262<br>(0.160) | 0.0641<br>(0.0898) |

[10] Desai & Andrist (2010) also argued that operationalisation of various dimensions of gender by measuring them at a regional or district level, has a clear advantage as the regional context plays a significant role in daughters' marriages.

| | | |
|---|---|---|
| Structure: | | |
| Index of white-collar job growth in region | 0.470<br>(1.214) | 0.237<br>(0.201) |
| Household transition | 1.726**<br>(0.800) | -1.427*<br>(0.733) |
| Female deficit in marriageable age cohort (bride deficit) | 0.327<br>(0.261) | |
| Prestige: | | |
| Occupation homogamy | -0.512**<br>(0.225) | -0.267<br>(0.311) |
| Cragg–Donald minimum eigenvalue statistic | 189.413 | 74.596 |
| Critical value (n =3, k = 3, b = 0.05) | 9.53 | 9.53 |
| N | 95,623 | 69,482 |

Note: Other controls include: intercept, age, age-square, 12 fractile classes of monthly per capita consumption expenditures, caste, religion, and region indicators. Omitted religion category is Hindu, omitted caste category is General, omitted region category is Central. ***1% level, **5% level, *10% level significance.
Abbreviation: 2SLS, two-stage least-squares regression; IV, instrumental variables.

For educational hypogamy in the rural sector (61st round), regional female white-collar workforce participation serves as the IV, capturing developmental idealism (Andrist et al., 2014; Allendorf & Pandian, 2016). This IV is significant, with a high Cragg-Donald F-statistic meeting Stock-Yogo criteria and strong correlation with the endogenous variable, confirming relevance.

**Table 4b**: 2SLS – IV model of the effect of education matching, stigma, structure, household transition, female deficit and MPCE classes on spousal age gap: urban sector

| | **61st round** | **68th round** |
|---|---|---|
| | all controls | all controls |
| Husband grad & wife primary/middle/secondary/higher secondary level of education | 0.231<br>(1.213) | -0.516<br>(0.987) |
| Husband and wife graduate | -1.180<br>(1.598) | -2.447<br>(2.311) |
| Husband grad & wife postgrad | -1.759<br>(4.659) | -0.880<br>(2.292) |
| Stigma: | | |
| Husband in white-collar job in blue-collar region | 0.269<br>(0.178) | 0.227<br>(0.181) |
| Structure: | | |
| Index of white-collar job growth in region | -3.279***<br>(0.453) | -0.0966<br>(0.282) |
| Household transition (non-transitional households in 61st round & transitional household in 68th round) | -0.185<br>(0.834) | -0.260<br>(0.924) |
| Female deficit in marriageable age cohort (bride deficit) | -2.293***<br>(0.752) | -0.496<br>(0.749) |
| Prestige: | | |
| Occupation homogamy | -0.564<br>(0.396) | -0.435<br>(0.536) |
| Cragg–Donald minimum eigenvalue statistic | 9.43 | 8.27 |
| Critical value (n =3, k = 3, b = 0.05) | 9.53 | 9.53 |
| N | 44,705 | 41,206 |

Note: Other controls include: intercept, age, age-square, 12 fractile classes of monthly per capita consumption expenditures, caste, religion, and region indicators. Omitted religion category is Hindu, omitted caste category is General, omitted region category is Central. ***1% level, **5% level, *10% level significance.
Abbreviation: 2SLS, two-stage least-squares regression; IV, instrumental variables.

In the 68th round (2011–12), demographic shifts, including weakening arranged marriage[11] and female deficits in marriageable age cohorts, necessitate a new IV for rural homogamy:

[11] In the past thirty years, arranged marriages in India have seen significant shifts. Andrist, Banerji, and Desai (2013) report a decline in traditional practices, while Allendorf & Pandian (2016) highlight evolving norms in

district-level[12] educational homogamy at the graduate level[13], combined with household per capita income and a regional bride squeeze index[14]. This IV is exogenous but shows weaker performance (Cragg-Donald F-statistic: 8.27, below Stock-Yogo critical value). The exactly identified model fails the Anderson-Rubin Wald test for joint significance of endogenous regressors, limiting confidence in instrument validity.

In the urban sector, district-level assortative mating is the IV for both rounds, with regional per capita income as a multiplier in the 61st round due to marriage migration amid regional poverty. In the 68th round, income is excluded, reflecting reduced poverty[15] and urban expansion[16]. Table 4 (a and b) reports 2SLS results: the 61st round IV is moderately valid (Cragg-Donald F-statistic: 9.43, near Stock-Yogo critical value of 9.53), but the standard Stock-Yogo test is inappropriate for just-identified models. Instead, the Hansen J test supports instrument validity, though the 68th round shows weaker validity, with an insignificant Anderson-Rubin Wald test. These limitations suggest that rural homogamy and hypogamy IVs are robust in the 61st round but face challenges in the 68th round due to demographic shifts, while urban IVs show

---

marital arrangements with increased involvement of women in choosing their spouses and a rise in pre-wedding meetings between engaged couples.

[12] Desai & Andrist (2010) contend that examining gender at a district level, rather than an individual level, offers the benefit of minimising endogeneity concerns, as age at marriage and various indicators of gender relations could be reciprocally related at the individual level.

[13] Alternatively, the proposed process-proxy—defined by local-level educational endogamy at the tertiary level, enhanced by household-level economic status, and the regional availability of females in the marriageable age cohort—serves as a potential precondition for cultural change. This notion is consistent with Guilmoto's (2009) insights on the prerequisites for a potentially reversible sex ratio transition in Asia. Nonetheless, the intellectual foundation for this cultural shift draws from the ground breaking framework of Ansley Coale. Coale identified specific triggers as "originators" for a sustained reduction in marital fertility in Europe. In his 1973 work, Coale highlighted three pivotal preconditions for a fertility transition:

1. A **willingness** to embrace a new, legitimate norm. This can be recognized in terms of ethical acceptance, like broad (spatial) endorsement of gender equality in education, or the diffusion of a shared standard. In our context, this is illustrated by the district-level educational congruence between partners (i.e., the diffusion and/or spatial acceptance of the gender equality in education).
2. The new behavior must offer an **economic benefit** to its adopters, driving them to align with or adapt to this new standard. For example, the societal or institutional acceptance of gender equality in education should have benefits that eclipse the costs—specifically, the price of abandoning age-old norms. Economic robustness, reflected in trends like rising per capita income or economic mobility, is crucial for fostering this **readiness**.
3. Participants must have the **ability** to capitalize on the innovation's advantages. For instance, a shortage of brides (representing a bottleneck scenario) could prompt men to compete for better prospect in marriage market as the risk of remaining unmarried in a marriage-oriented society could be trivial. Conversely, in a gender-biased society (with lack of avenues of mobility), the potential for marital mobility might encourage women to strategically position themselves—using education as a signal of resilience against redundancy—in the market.

Our instrument can, therefore, potentially capture the framework for cultural change, making it apt for modeling the unobserved effect of educational homogamy.

[14] Demographic research suggests that men residing in remote rural areas with lower social status and limited social resources may face challenges in the marriage market, as they are less appealing to women who prefer partners with higher socioeconomic status (Greenhalgh & Winkler, 2005; Y. Li et al., 2010; Xueyan, Yang & Li, Shuzhuo & Attané, Isabelle & Feldman, Marcus. (2016). On the Relationship Between the Marriage Squeeze and the Quality of Life of Rural Men in China. American Journal of Men's Health. 11. 10.1177/1557988316681220.

[15] According to the Economic Survey 2013-14, published by the Ministry of Finance, Government of India, the poverty ratio in India (derived from the Monthly Per Capita Expenditure (MPCE) thresholds of Rs. 816 for rural areas and Rs. 1000 for urban areas in 2011-12) experienced a substantial decline from 37.2% in 2004-05 to 21.9% in 2011-12.

[16] The increase in urban population during 2001-2011 was not only the highest registered thus far but also marginally exceeded the corresponding increase in rural population, Census GOI, 2011.

moderate validity in the 61st round but sensitivity to changing economic and social conditions in the 68th round.

## 6. Conclusion and Discussion

Using the waves of NSS datasets (50th, 61st, and 68th) of the employment and unemployment, this study elucidates the persistent and evolving determinants of spousal age gap in India, contextualized within the broader framework of socioeconomic development, institutional transformation, and evolving gender norms. Our analysis demonstrates that while traditional age hypergamy remains deeply entrenched, its magnitude is shaped significantly by the interplay of the intersection of educational assortative mating patterns, household economic status, and localized labor market structures.

Our findings reveal a nonlinear relationship between spousal educational differences and age gaps. Marriages characterized by male educational advantage—particularly where husbands possess graduate-level education and wives hold comparatively lower qualifications—exhibit substantively larger age gaps across rural and urban India, underscoring ingrained gender hierarchies that confer social and economic leverage to men. Conversely, homogamous and hypogamous pairings, particularly where wives possess advanced education, are associated with narrower age gaps, suggesting a gradual erosion of traditional constraints as women's socio-economic agency strengthens.

The influence of economic stratification, operationalized via granular consumption expenditure fractiles, further substantiates this complexity. In rural contexts, increments in household wealth are positively correlated with increased age gaps, reflecting the entrenched social capital men wield in marriage markets that valorise economic resources. In contrast, urban data reveal a threshold effect wherein the relationship attenuates or reverses at higher wealth levels, indicative of progressive cultural shifts and emergent egalitarian preferences among the affluent.

Moreover, social norms and institutional constraints play a mediating role. The stigma index reveals that entrenched gender role expectations—particularly in regions where male white-collar employment dominates amidst blue-collar economic environments—exacerbate age hypergamy by constraining women's labor market participation and autonomy. Meanwhile, occupational homogamy emerges as a salient proxy for nascent gender egalitarianism within households, robustly associated with reductions in spousal age gaps and highlighting the interplay between professional parity and intra-household negotiation power.

Structural transformations in household arrangements further modulate these effects. The proliferation of transitional household types correlates with diminished age gaps, highlighting avenues through which evolving family forms erode patriarchal dominance and bolster female agency. The bride deficit, reflecting skewed sex ratios and demographic imbalances in the marriageable age cohort, initially exacerbates age hypergamy by intensifying male-favourable bargaining conditions but appears to engender adaptive social responses over time that improve women's relative standing in the marital market.

These findings contribute to the growing literature on the demographic-economic nexus by highlighting how education, economic development, and gender norms co-evolve to shape marriage market outcomes. From a policy perspective, strategies that expand female educational attainment beyond critical thresholds—thereby increasing their opportunity costs and bargaining power—are paramount. Simultaneously, addressing labor market discrimination and stigma, particularly in manufacturing and traditionally male-dominated sectors, can dismantle barriers to women's economic participation. Furthermore, social policies facilitating flexible and supportive household structures may accelerate cultural shifts towards more equitable marital arrangements.

While this analysis employs rigorous econometric techniques and extensive representative data, limitations persist, including reliance on cross-sectional household consumption as a proxy for income and the absence of direct measures on individual preferences and intra-family decision-making. Additionally, the persistence of arranged marriage systems and cultural heterogeneity[17] (as the NSS does not provide this information) implies that unobserved factors may continue to influence these dynamics. Future research employing longitudinal designs, integrated with qualitative insights into shifting social norms and evolving marriage market mechanisms—especially in emerging urban contexts—is essential to unpack the complex feedback loops shaping spousal age gaps. Understanding how technological, demographic, and institutional changes intersect to influence gender relations remains a critical frontier.

By advancing understanding of the multifactorial drivers of age hypergamy, this study contributes to an enriched discourse on gender relations and social stratification in India's ongoing socioeconomic transition. Policies fostering female empowerment and inclusive economic participation hold promise for reshaping marital patterns toward greater equity and social inclusiveness.

[17] The primary determinants of the age at marriage in India are region and its geographical manifestations, culture, the respondent's level of education, caste, religion, wealth, mass media exposure, etc. (Jejeebhoy & Sathar, 2001; Singh, 2005; Cislaghi et al., 2020), most of which have already been considered in our model and are, therefore, endogenous.

**Appendix 1**:

**Table 1: Spousal age gap:** Census of India, 1991, 2001 & 2011

| | **Rural** | | | **Urban** | | |
|---|---|---|---|---|---|---|
| Area Name | 1991 | 2001 | 2011 | 1991 | 2001 | 2011 |
| India | 4.7 | 4.6 | 4.4 | 4.9 | 4.7 | 4.6 |
| Jammu & Kashmir | | 3.7 | 3.1 | | 3.2 | 2.2 |
| Himachal Pradesh | 4.2 | 4.2 | 4.3 | 4 | 3.3 | 3.2 |
| Punjab | 3.3 | 3.2 | 3.1 | 3.5 | 3.1 | 3.2 |
| Uttarakhand | | 4.2 | 4.1 | | 4 | 3.9 |
| Haryana | 4 | 4 | 3.9 | 3.9 | 3.7 | 3.5 |
| Delhi | 3.8 | 3.9 | 3.8 | 3.6 | 3.7 | 3.6 |
| Rajasthan | 4.1 | 3.7 | 3.4 | 4.2 | 3.9 | 3.6 |
| Uttar Pradesh | 4 | 3.8 | 3.5 | 4.2 | 4 | 3.9 |
| Bihar | 4.7 | 4.5 | 4.2 | 5.2 | 5.3 | 5 |
| Arunachal Pradesh | 5 | 4.1 | 3.4 | 5.3 | 4.5 | 3.2 |
| Nagaland | 3.9 | 3.1 | 2.9 | 4.8 | 4.2 | 3 |
| Manipur | 3.4 | 3.1 | 2.7 | 3.6 | 2.9 | 2.9 |
| Mizoram | 4.4 | 3.9 | 3.1 | 4.2 | 3.2 | 2.2 |
| Tripura | 6 | 5.6 | 5.1 | 6.5 | 6.2 | 6.4 |
| Meghalaya | 4.4 | 4.2 | 3.6 | 4 | 3.6 | 3 |
| Assam | 6.1 | 5.7 | 5.5 | 6.3 | 6.1 | 6 |
| West Bengal | 6.4 | 6 | 5.8 | 6 | 6.1 | 6 |
| Jharkhand | | 4.5 | 4.2 | | 5 | 4.8 |
| Orissa | 4.7 | 4.7 | 4.8 | 5.5 | 5.5 | 5.5 |
| Chhattisgarh | | 3.3 | 3.2 | | 4.5 | 4.1 |
| Madhya Pradesh | 3.9 | 3.7 | 3.5 | 4.6 | 4.4 | 4.2 |
| Gujarat | 3.3 | 3.3 | 3.3 | 3.8 | 3.8 | 3.7 |
| Maharashtra | 5.2 | 5 | 4.8 | 4.9 | 4.6 | 4.4 |
| Andhra Pradesh | 5.3 | 4.9 | 4.8 | 5.3 | 5 | 4.8 |
| Karnataka | 6.1 | 5.9 | 5.9 | 6.2 | 5.9 | 5.6 |
| Goa | 5.3 | 5 | 5.3 | 5.1 | 5 | 5.1 |
| Kerala | 5.4 | 6.2 | 6.8 | 5.8 | 6.5 | 7.2 |
| Tamil Nadu | 5.4 | 5.4 | 5.7 | 5.7 | 5.5 | 5.4 |

*Source*: Author computation from different Census rounds, 1991-2011

**Appendix 2**:

**Table 2: Spousal age gap: Census of India,** NSS 50th (1993-94), 61st (2004-05) and 68th (2011-12) rounds, Rural and Urban India

| | **Rural** | | | **Urban** | | |
|---|---|---|---|---|---|---|
| **State Name** | 50th | 61st | 68th | 50th | 61st | 68th |
| India | 5.1 | 4.9 | 4.8 | 5.6 | 5.3 | 5.1 |
| Jammu & Kashmir | 6.4 | 3.9 | 4.2 | 5.9 | 3.7 | 4.3 |
| Himachal Pradesh | 6.0 | 5.4 | 5.0 | 6.1 | 5.3 | 5.0 |
| Punjab | 3.9 | 3.5 | 3.5 | 4.4 | 3.9 | 3.7 |
| Uttarakhand | | 5.1 | 4.5 | | 4.9 | 4.7 |
| Haryana | 3.9 | 3.8 | 3.5 | 4.5 | 4.1 | 3.6 |
| Delhi | 4.5 | 3.5 | 3.0 | 4.7 | 4.3 | 3.9 |
| Rajasthan | 3.7 | 3.5 | 3.4 | 4.2 | 4.0 | 3.8 |
| Uttar Pradesh | 3.7 | 3.4 | 3.4 | 4.3 | 4.0 | 4.0 |
| Bihar | **4.4** | **4.4** | **4.4** | 5.1 | 4.7 | 4.7 |
| West Bengal | 7.0 | 6.8 | 6.9 | 7.5 | 7.2 | 7.0 |
| Jharkhand | | 5.0 | 5.0 | | 5.7 | 5.2 |
| Orissa | 5.6 | 5.5 | 5.2 | 6.3 | 6.0 | 5.8 |
| Chhattisgarh | | 3.9 | 3.8 | | 4.8 | 4.7 |
| Madhya Pradesh | 3.9 | 3.6 | 3.7 | 5.0 | 4.4 | 4.4 |
| Gujrat | 3.6 | 3.2 | 3.0 | 4.3 | 3.8 | 3.7 |
| Maharashtra | 5.9 | 5.5 | 5.1 | 5.9 | 5.6 | 5.2 |
| Andhra Pradesh | 6.1 | 5.5 | 5.8 | 5.9 | 5.8 | 5.6 |
| Karnataka | 6.8 | 6.2 | 6.1 | 7.1 | 6.5 | 6.2 |
| Kerala | 6.5 | 6.6 | 6.7 | 7.1 | 7.0 | 6.6 |
| Tamil Nadu | 6.7 | 6.4 | 6.2 | 6.9 | 6.1 | 6.0 |
| North east | 6.7 | 6.3 | 5.8 | 6.9 | 6.3 | 6.1 |

*NOTES*: Estimates of means obtained using survey weights.

*SOURCE*: NSS 50th, 55th, 61st & 68th round Employment and Unemployment Survey.

**Appendix 3**: **Coefficient plot of Income class on age hypergamy** (full model)

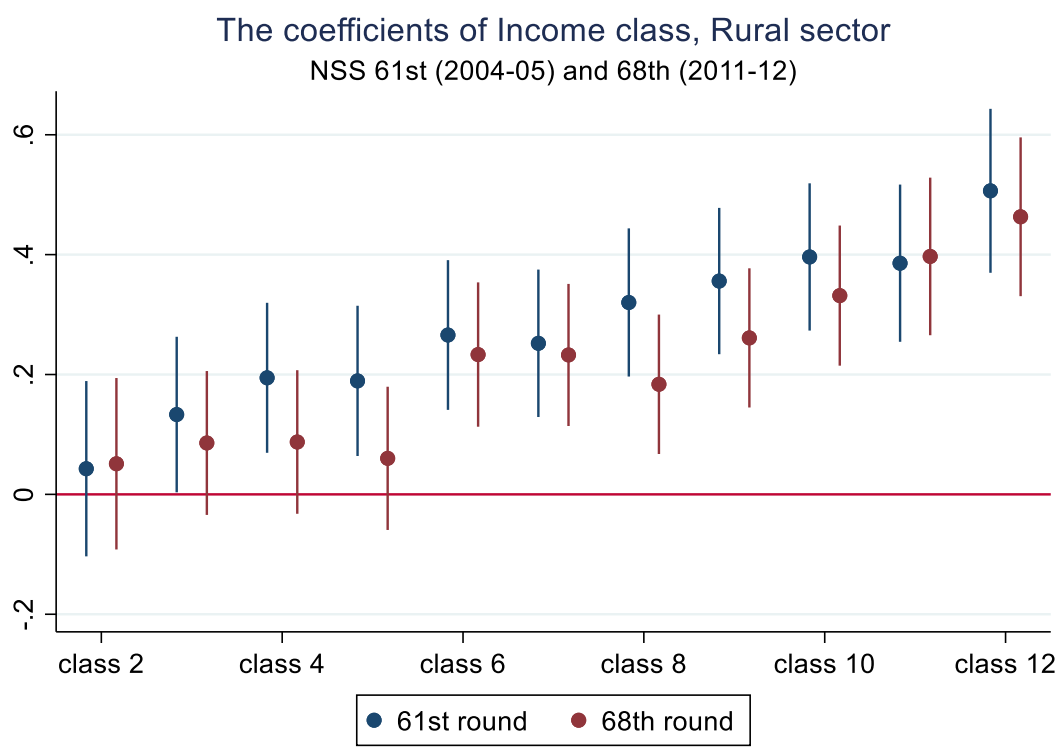

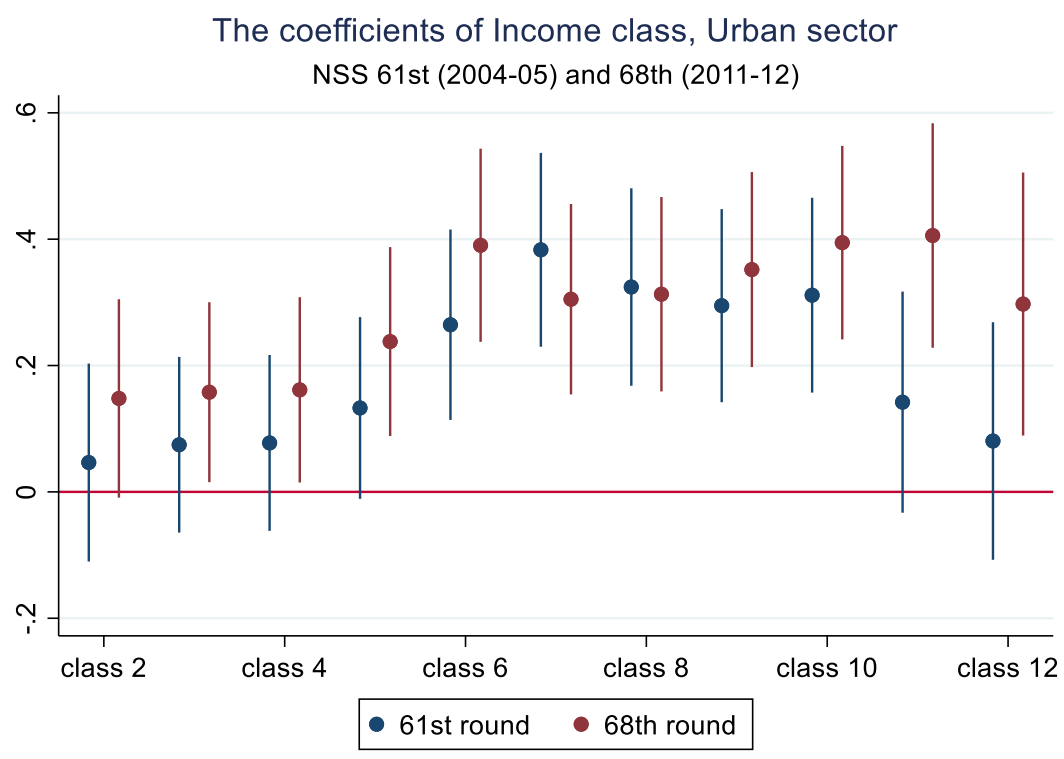

## **Appendix 4(a)**: Rural Sector

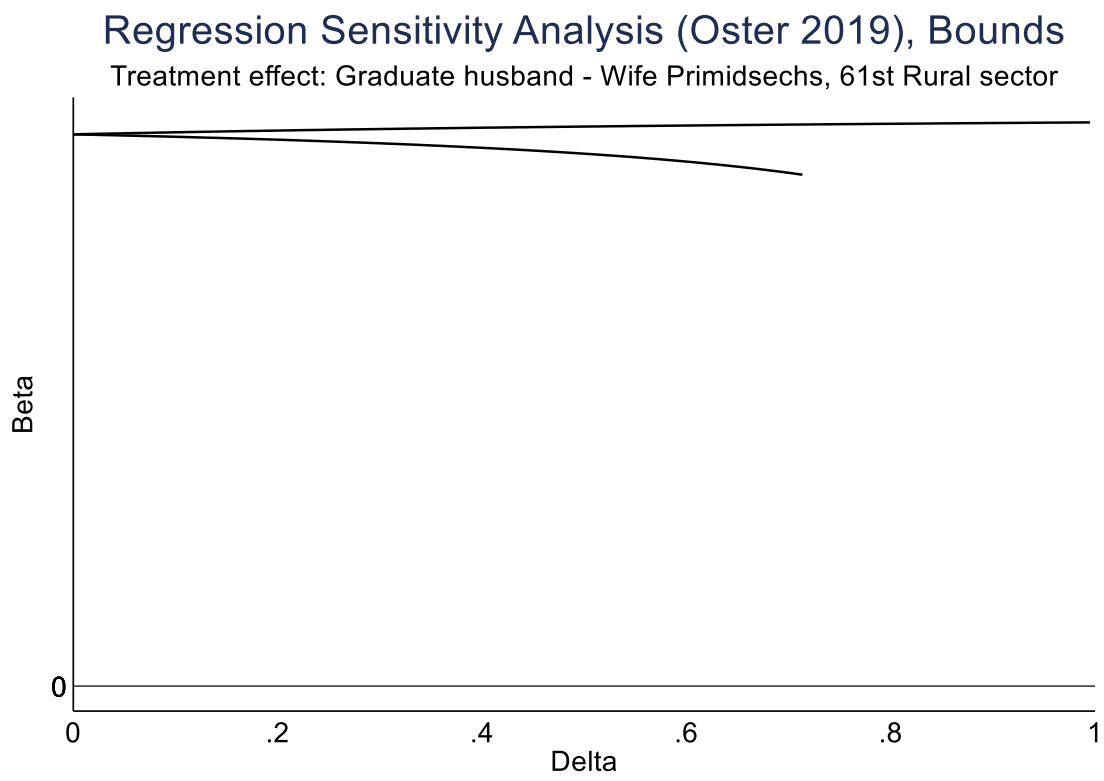

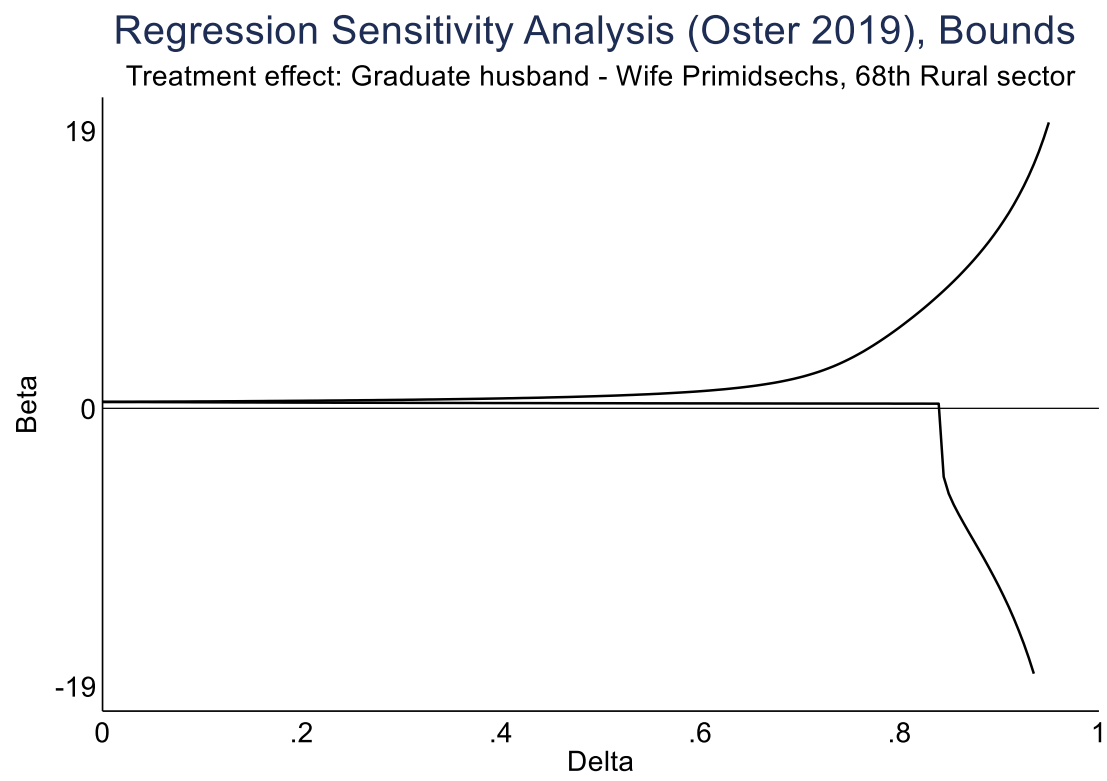

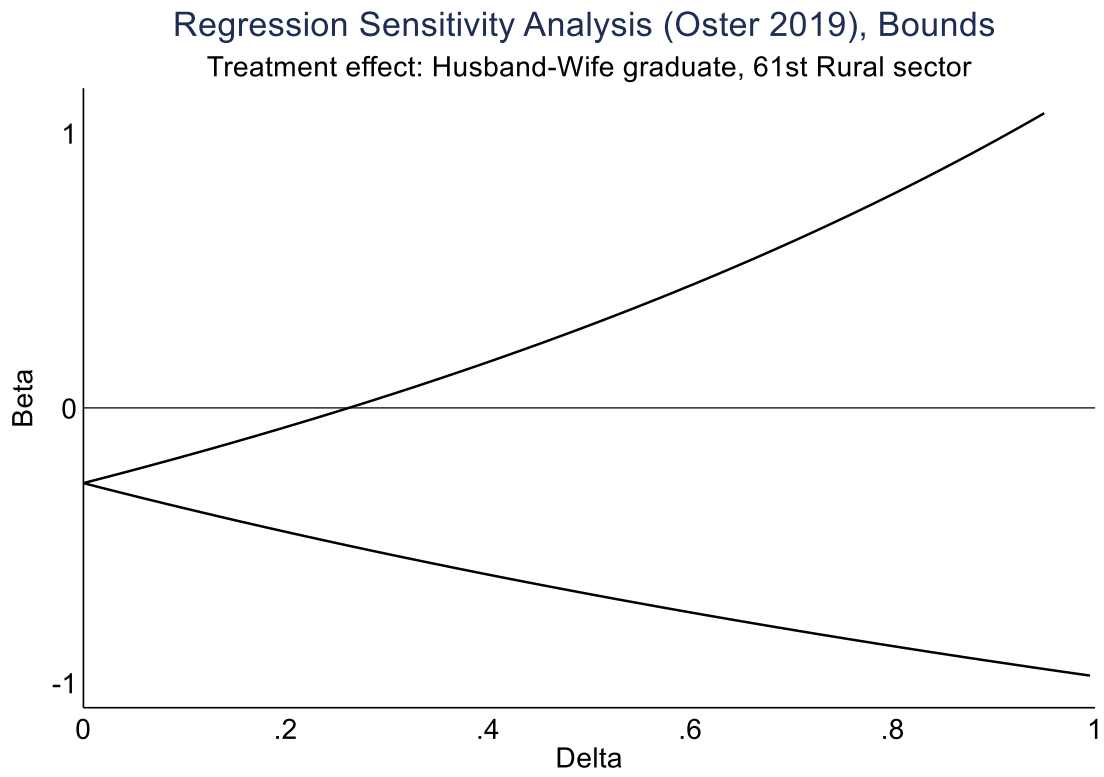

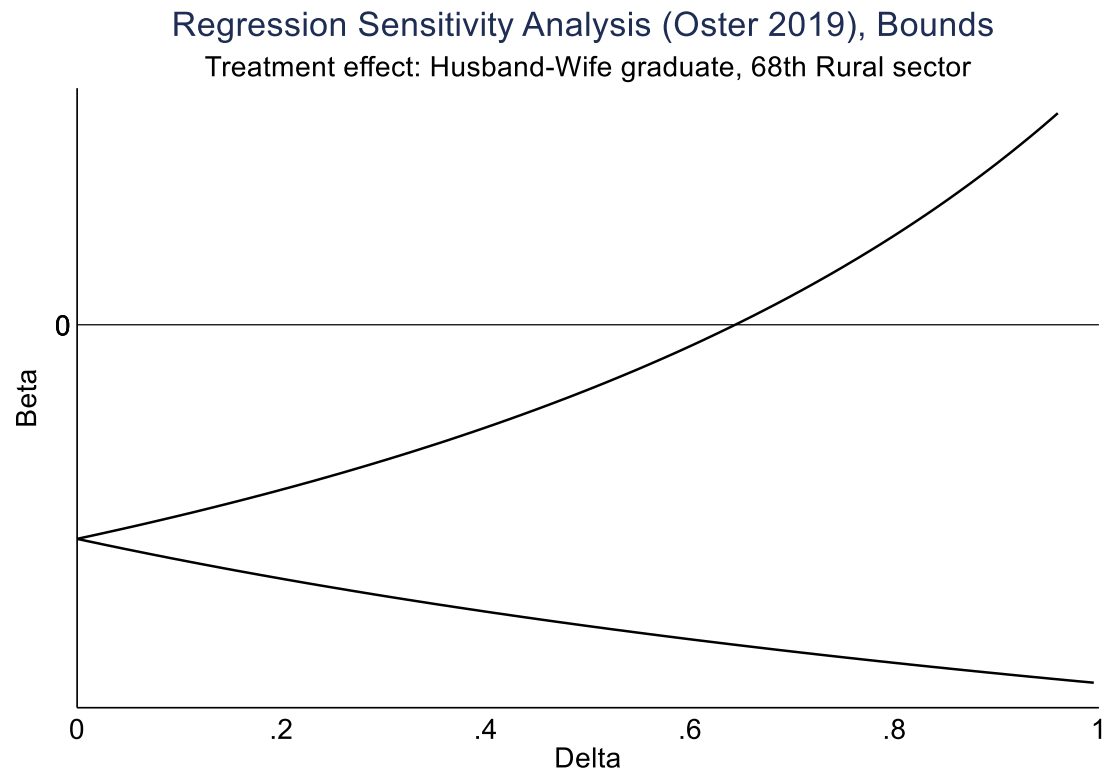

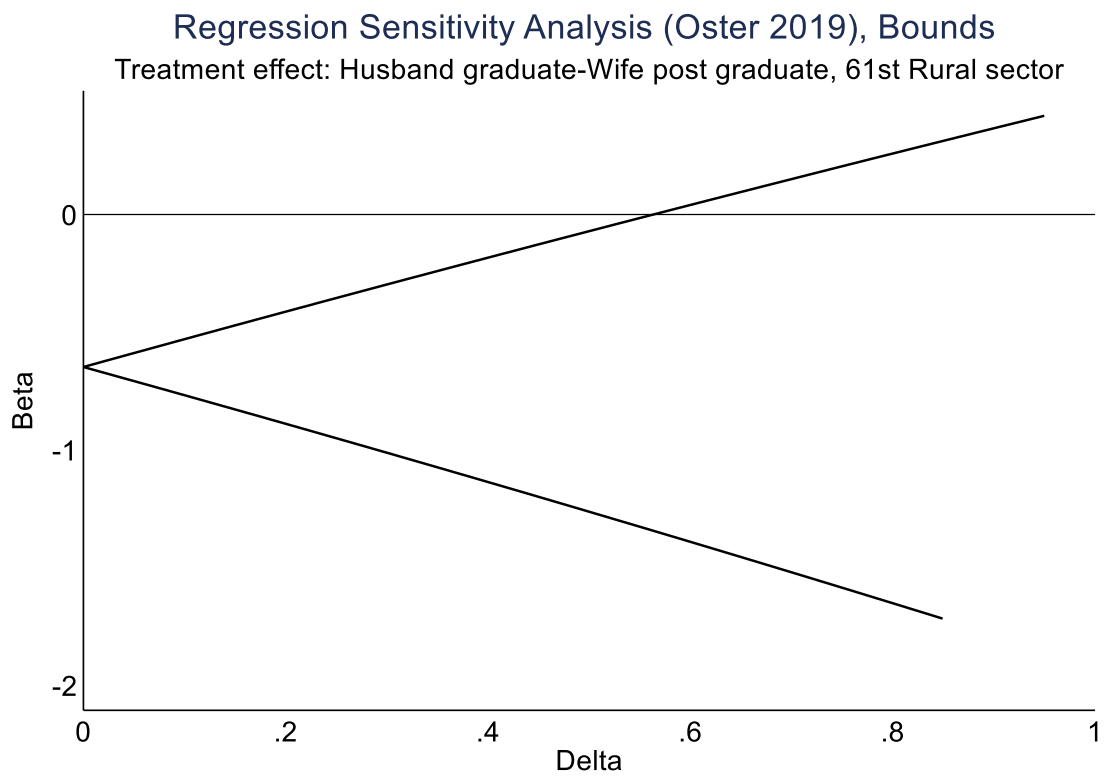

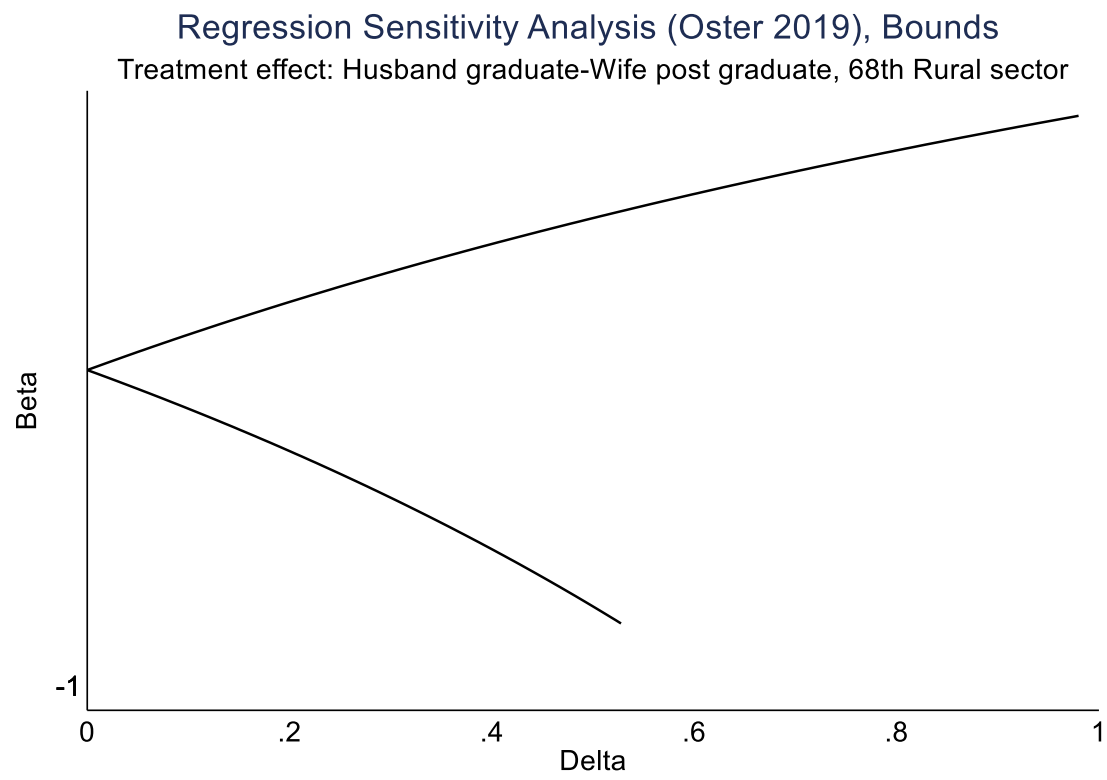

**Appendix 4(b)**: Urban sector

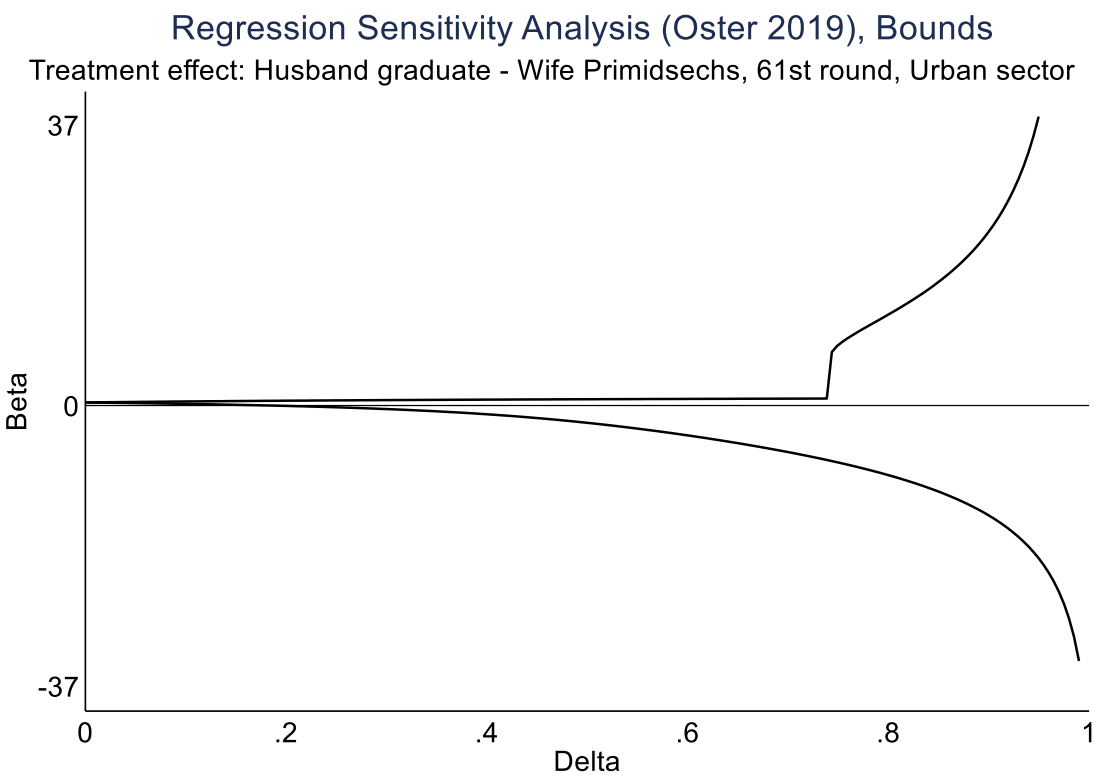

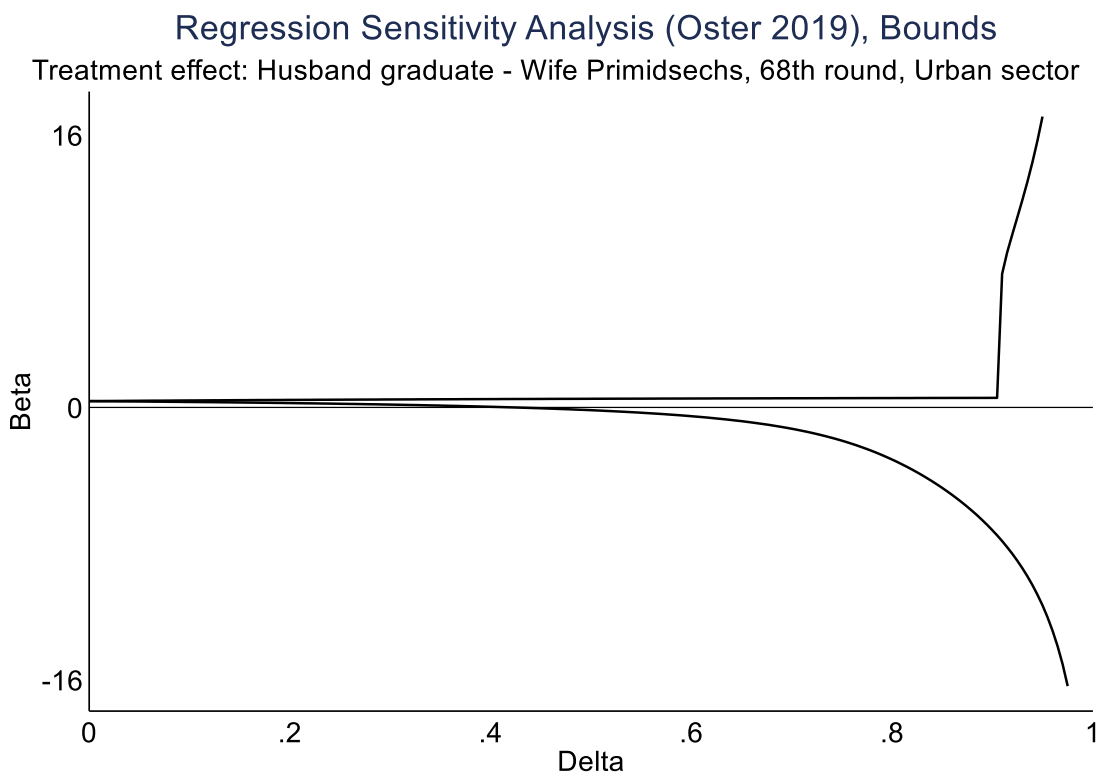

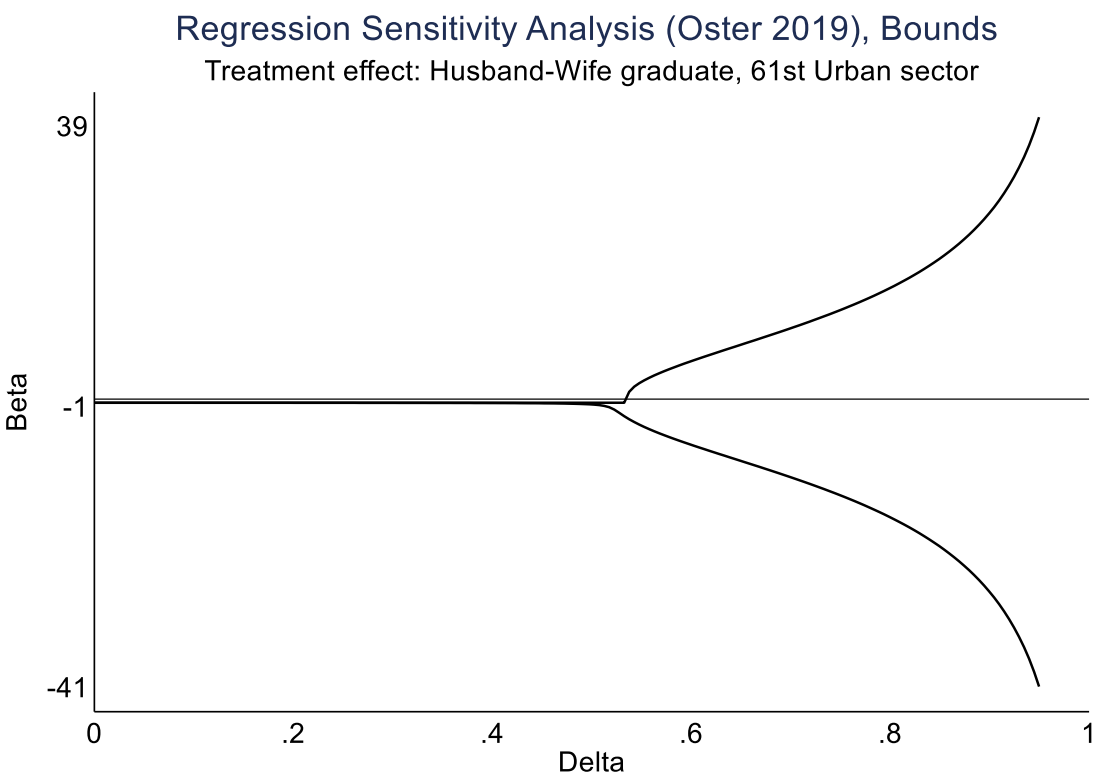

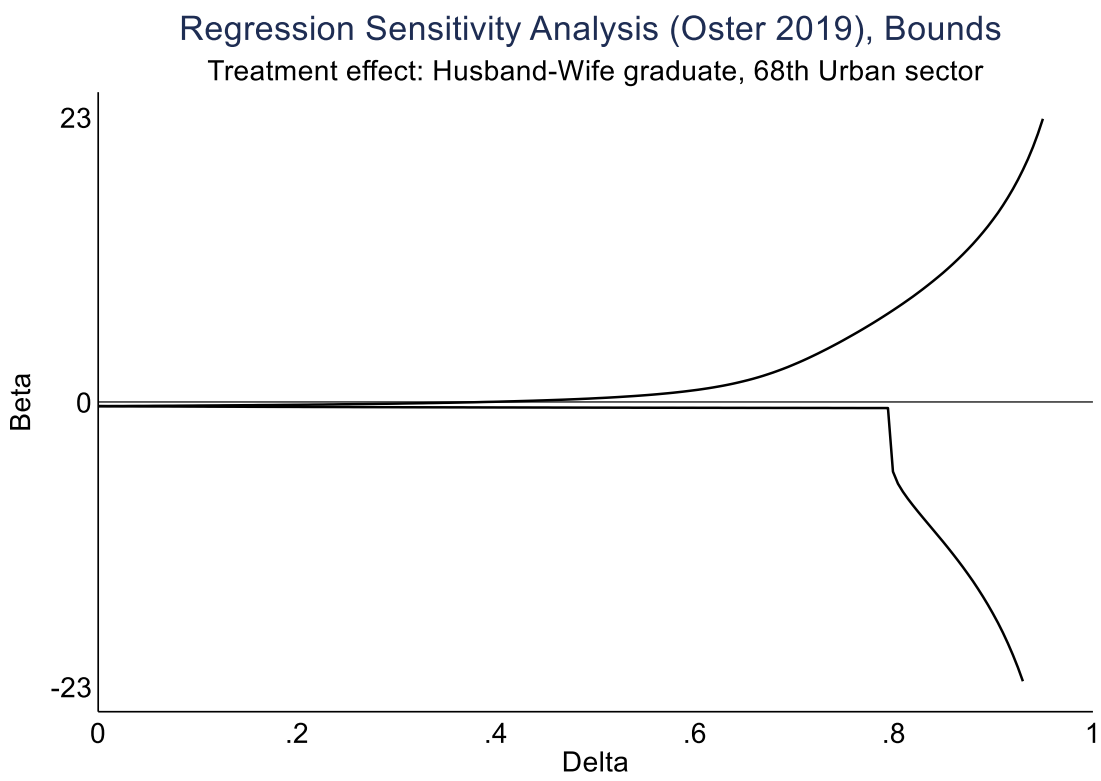

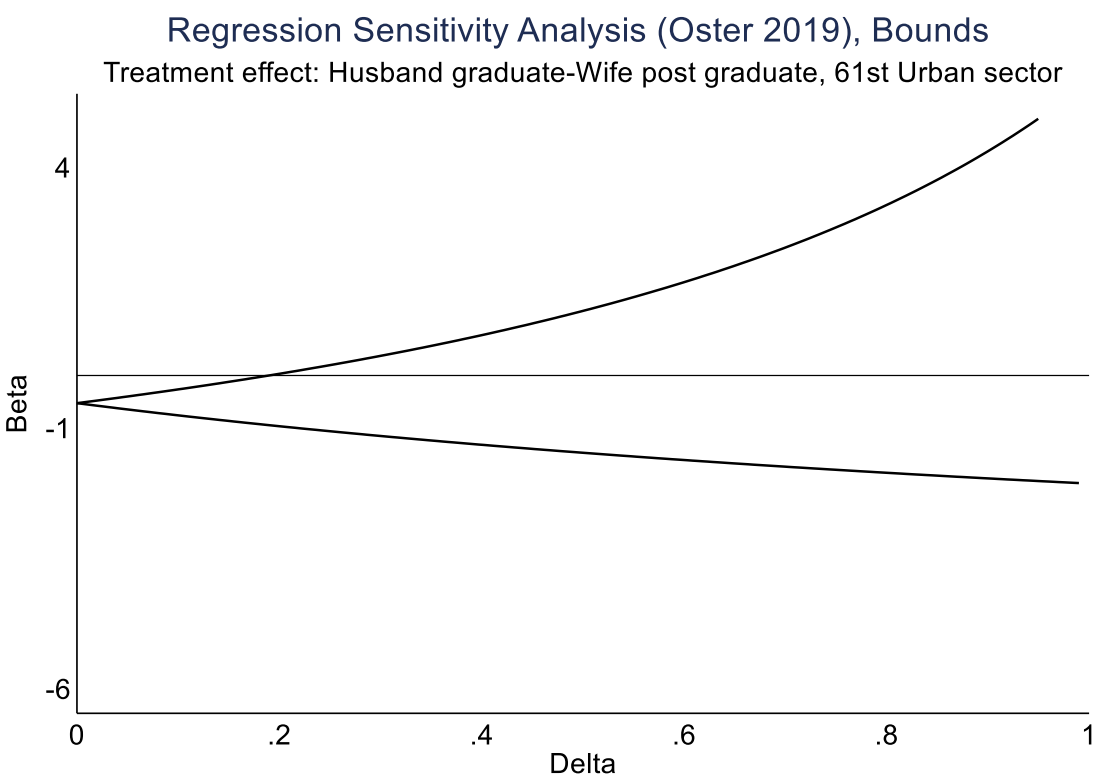

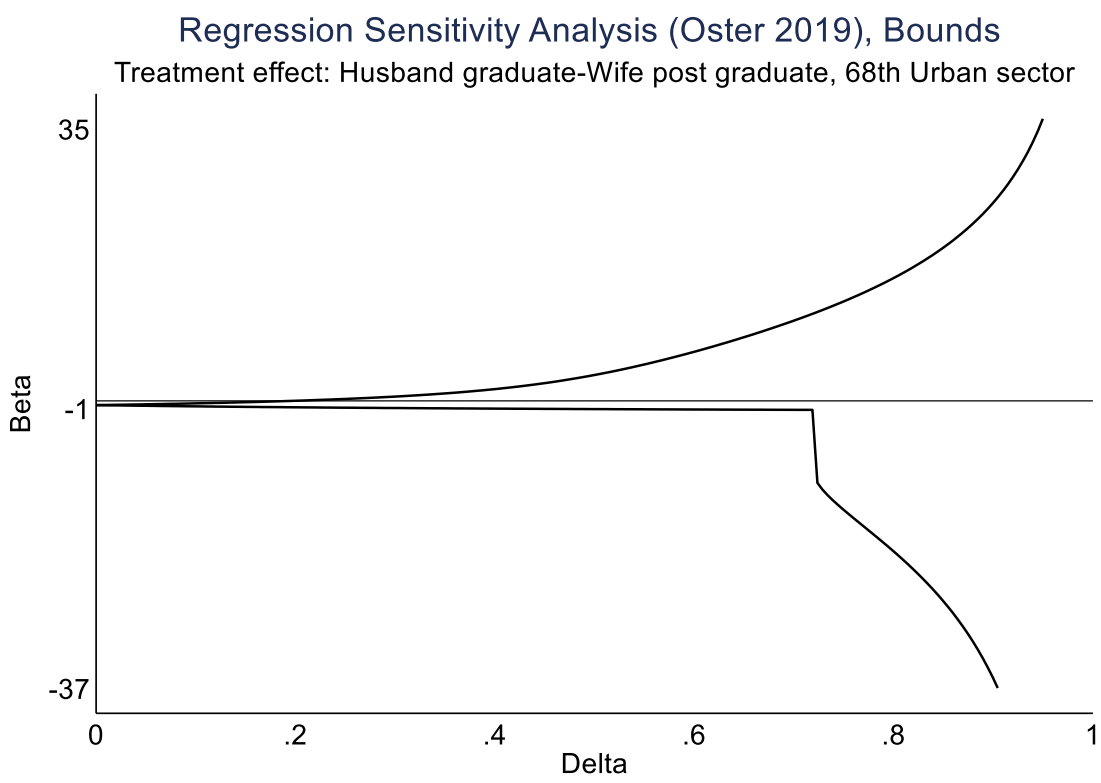